\pdfoutput=1
\documentclass[epjST]{svjour}
\usepackage{graphicx,url}
\usepackage[latin1]{inputenc}
\usepackage{listings}
\usepackage{graphicx}%
\usepackage{multirow}%
\usepackage{mathrsfs}%
\usepackage[title]{appendix}%
\usepackage{xcolor}%
\usepackage{textcomp}%
\usepackage{manyfoot}%
\usepackage{booktabs}%
\usepackage{algorithm}%
\usepackage{algorithmicx}%
\usepackage{algpseudocode}%
\usepackage{listings}%

\usepackage[font={small, sl}]{caption}

\usepackage{xspace}
\usepackage{color}
\usepackage{epsfig}     
\graphicspath{{.}{./figures/}}

\usepackage{keycommand} 
 \usepackage{hyperref}
  \usepackage{wrapfig}

\def\bR{\begin{color}{red}}  
\def\bB{\begin{color}{blue}}
\def\bM{\begin{color}{black}}  
\def\bC{\begin{color}{cyan}}
\def\bW{\begin{color}{white}}
\def\bBl{\begin{color}{black}}
\def\bG{\begin{color}{green}}
\def\bY{\begin{color}{yellow}}
\def\e{\end{color}\xspace}
\newcommand{\bit}{\begin{itemize}}
\newcommand{\eit}{\end{itemize}\par\noindent}
\newcommand{\ben}{\begin{enumerate}}
\newcommand{\een}{\end{enumerate}\par\noindent}
\newcommand{\beq}{\begin{equation}}
\newcommand{\eeq}{\end{equation}\par\noindent}
\newcommand{\beqa}{\begin{eqnarray*}}
\newcommand{\eeqa}{\end{eqnarray*}\par\noindent}
\newcommand{\beqn}{\begin{eqnarray}}
\newcommand{\eeqn}{\end{eqnarray}\par\noindent}

\begin{document}

\title{Quantum Concept Music from Categorical Quantum Mechanics:
Musical Incarnation of Entanglement}

\author{
Rakhat-Bi Abdyssagin\inst{1, 7}\fnmsep\thanks{\email{rahatbiabd@gmail.com}}
\and
Bob Coecke\inst{2, 3, 4, 5, 6}\fnmsep\thanks{\email{bob@rel-int.ai}}
} 

\institute{
Moth, London, UK 
\and
Quantum Brain Art Ltd, Oxford, UK
\and
Relational Intelligence Ltd, Oxford, UK
\and
Quantum Entertainment Berlin, Germany 
\and
Wolfson College, University of Oxford, UK
\and
Perimeter Institute, Waterloo, Ontario, Canada
\and 
Kulyash Bayseyitova Kazakh National University of Arts, Kazakhstan
}
 
 \abstract{We initiate the development of a new language and theory for quantum music, to which we refer as Quantum Concept Music (QCM). This new music formalism is based on Categorical Quantum Mechanics (CQM), and more specifically, its diagrammatic incarnation Quantum Picturalism (QPict),  which is heavily grounded on ZX-calculus. In fact, QCM is naturally inherited from CQM/QPict, \bM given that the category-theoretic nature of the latter allows for varying underlying models, to process theories of real-world phenomena like music\e. \\
 Adopting a relational view of quantum theory--which is implicit in CQM/QPict--at QCM's heart is the explicit notational representation of relations that exist within and between the key concepts of music composition, performance, and automation. QCM also enables one to directly translate quantum phenomena into music compositions in a both intuitively obvious, rigorous, and mechanical manner.\\
 Following this pattern, we propose a conceptual way of writing scores for musicians interacting like a Bell-pair under measurement, and outline how it can be live performed. While most of the Western classical music notation has heavily relied on linear representation of music--which does not always adequately capture the \bM relational features\e of music--our approach is distinct by \bM making the relational dimensions of music first-class citizens, augmented with quantum-native features\e. \\
 This quantum-based technique not only influences the music at the profound level of composition, but also has a direct impact on a live performance, and also provides a new template for automating music, e.g.,~in the context of AI-generation. \\
Altogether, we initiate the creation of new music formalism that is powerful and efficient in capturing the  interactive nature of music, both in terms of internal and external interactions, and goes beyond the boundaries of Western classical music notation, which allows to use it in many different genres and orchestrations.}

\maketitle


\section{Introduction}

When passing from classical to quantum physics we change the language in which we describe nature. In first instance, mostly because of humanity's  inertia, this language is as close as possible to things we are familiar with from classical physics. As such, Schr\"odinger introduced wave mechanics \cite{schrodinger1926quantisierung}. However, later, when John von Neumann sought to gain some profound understanding of the new theory's implications, it became a formalism in its own right, namely the Hilbert space based quantum formalism \cite{vN}--which interestingly, von Neumann later denounced \cite{Redei1}.  

Categorical Quantum Mechanics (CQM) \cite{AC1,CD2,CPaqPav,CDKZ} and its diagrammatic incarnation called Quantum Picturalism (QPict) \cite{Kindergarten,ContPhys,CKbook,QiP} takes this a step further, by embracing from the very start that what has been identified by Schr\"odinger  as ``not one but the characteristic trait of quantum theory", namely how systems compose \cite{SchrodingerComp}. This `relational' focus also goes hand-in-hand with a view on nature in terms of `processes' \cite{JTF}, that is, following Heraclitus rather than Parmenides \cite{VerelstCoecke1999}. The new  formalism makes it much more intuitively clear how quantum theory is different from classical theory, and what it enables.  Hence, the  formalism is a direct witness of the new theory and its implications, and not just a vehicle for expressing it.   This is not merely a convenience or ideology, as witnessed by the successes of CQM/QPict in a wide spectrum ranging from complex problems in quantum engineering \cite{coecke2021kindergarden,van2020zx,khesin2023graphical,Bombin2024UnifyingFlavors}, studies in quantum foundations \cite{CES,MiriamSpek,CDKZ,StefanoWill,pinzani2019categorical},  
to quantum education.   Within the context of this paper the latter is particularly important: QPict has proven to have the potential to revolutionise quantum education by making quantum accessible at high school level and maybe even earlier \cite{Coecke2025HighSchoolers,dundar2025making}.

\begin{figure}[h]
   \centerline{\includegraphics[height=120pt]{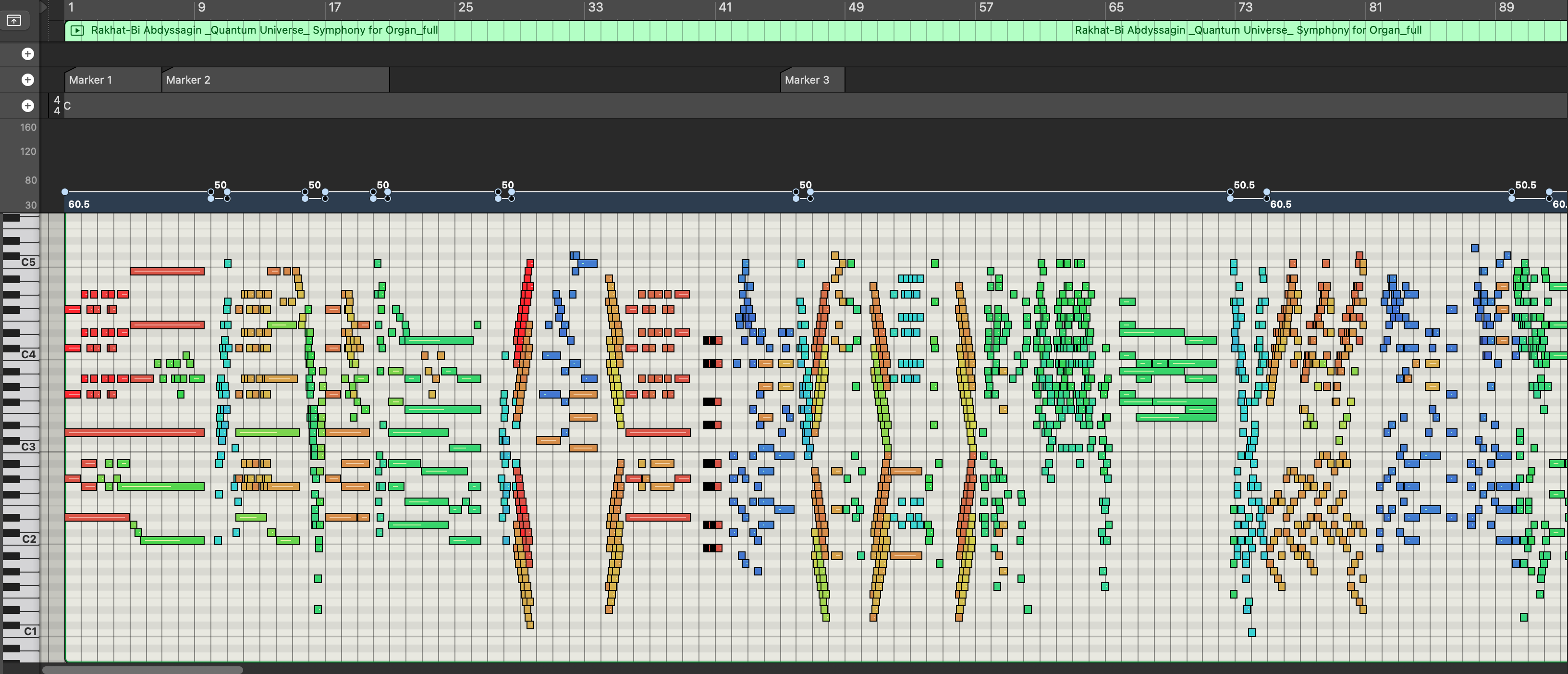}} 
   \caption{\bM Piano roll representation of Abdyssagin's recent ``Quantum Universe" Symphony for Organ, Quantum Guitar and Electronics (2025) in Logic Pro, most interestingly, exposes Dirac's ket- and bra-shapes.\e}  
    \label{QU}
\end{figure}

Of course, traditional music notation also comes with implicit bias to what traditional Western music was supposed to be, for example, with the visual `distances' between elements (e.g.,~noteheads, or even bars/measures) not matching the real proportions in music (e.g.,~temporal-durational distances, etc.). While suitable for the genres of Western classical music, it is not necessarily always adequate for thinking about music in a broader sense. \bM Piano-roll notation in software like Logic Pro is in that sense more faithful--See Fig.~\ref{QU}. \e

There are many more issues with it of course. For example, starting from the twentieth century, many distinguished composers have been experimenting with new forms and methods of music notation, because the `traditional' music notation proved to be inadequate to represent new music. New ideas need new formalism, thus, several different directions and modifications of notation appeared. Some of the striking and experimental but efficient and lucid examples of modified music notation within the mainstream of Western avant-garde music include works by Stockhausen 
\cite{Stockhausen1,Stockhausen2}, 
Cage 
\cite{Cage1,Cage2}, 
Murail \cite{Murail}, Grisey \cite{Grisey}, Lutoslawski \cite{Luto1}, Maderna \cite{Maderna}, Lachenmann 
\cite{Lach1,Lach2}, 
Ferneyhough \cite{BFern}, 
and many others. Most of the modifications to the notation that were introduced by avant-garde music composers aimed at capturing the new sonoristic effects and extended methods of playing instruments. Some other modifications changed the approach to time in music. Nevertheless, in some cases each composer or composition school had its own system of notation, and thus overall when the new extended instrumental techniques emerged, it was not easy to find consistency in notation. Even now this largely depends on the individual preferences, and it became almost customary that composers preface their scores with legend/performance notes where every non-classical technique is being defined and described to facilitate the performance and (in certain cases) the perception of the score.

When passing to quantum music \cite{miranda2022quantum} it makes perfect sense to have a radical change in music language and theory, which we conceive of together under the umbrella of `formalism',  just as in the case of CQM/QPict. Retaining some of the traditional features, the  formalism should be visual, but in a way that it tells the story about what the music is genuinely about, in particular its relational content, and not just something that happens to suit e.g.,~playing on a piano keyboard. 

The obvious way to do so is for this passage to stay close to the one that resulted into QPict,  given its fundamental relational and process-based foundation.  Unlike other methods of notation, ours is not a mere modification or alteration, but a completely new, innovative and quantum-driven formalism, that also will make music notation more accessible.  Given the importance of the notion of ``composition" within music, this seems like a match made in heaven.  And this compositional structure should follow the important transition of this concept from ``classical music" to ``quantum music'', which again can be enabled by QPict. Do note the deliberate ambiguous use of ``composition" and ``classical" here.

As an alternative formalism for quantum theory that brings relationships to the forefront, QPict is based on the following principle:
\smallskip\par\noindent
\emph{Whenever there  \bM is a  relationship\e between entities in the  described phenomenon, then these connections are also visible in the notation.} 
\smallskip\par\noindent
\bM However, as quantum theory includes non-classical phenomena like complementarity, measurement non-determinism, and collapse, these relations are of a very different nature than typical classical constraints. Indeed, quantum theory is not merely a theory of constraints. \e

Translating this to music, our proposed formalism provides a vehicle that directly exhibits what actually is being thought of, composed or performed, as well as a direct bridge between musical composition and performance.  All of this carries evidently over to music automation, and  all of this  while at the same time providing a direct bridge with  the quantum world, \bM as we explain in the following section\e.

In this paper we assume familiarity with basic quantum theory, and basic ZX-calculus \cite{coecke2023basic}. For our notation, we will borrow the spiders of QPict \cite{CKbook,QiP}. 
\bM Fundamental quantum features arise by having spiders of two colours, corresponding to complementary observables, and phases, which generate the Block-sphere geometry \bM arising from complex numbers \cite{CD1,CD2}. Quantum measurements and mixed states are automatically induced by the spiders \cite{CPaqPav}\e.  These spiders also have the tensor-structure built-in.  We elaborate on this in the next section. 

\begin{figure}[h!]
   \centerline{\includegraphics[height=120pt]{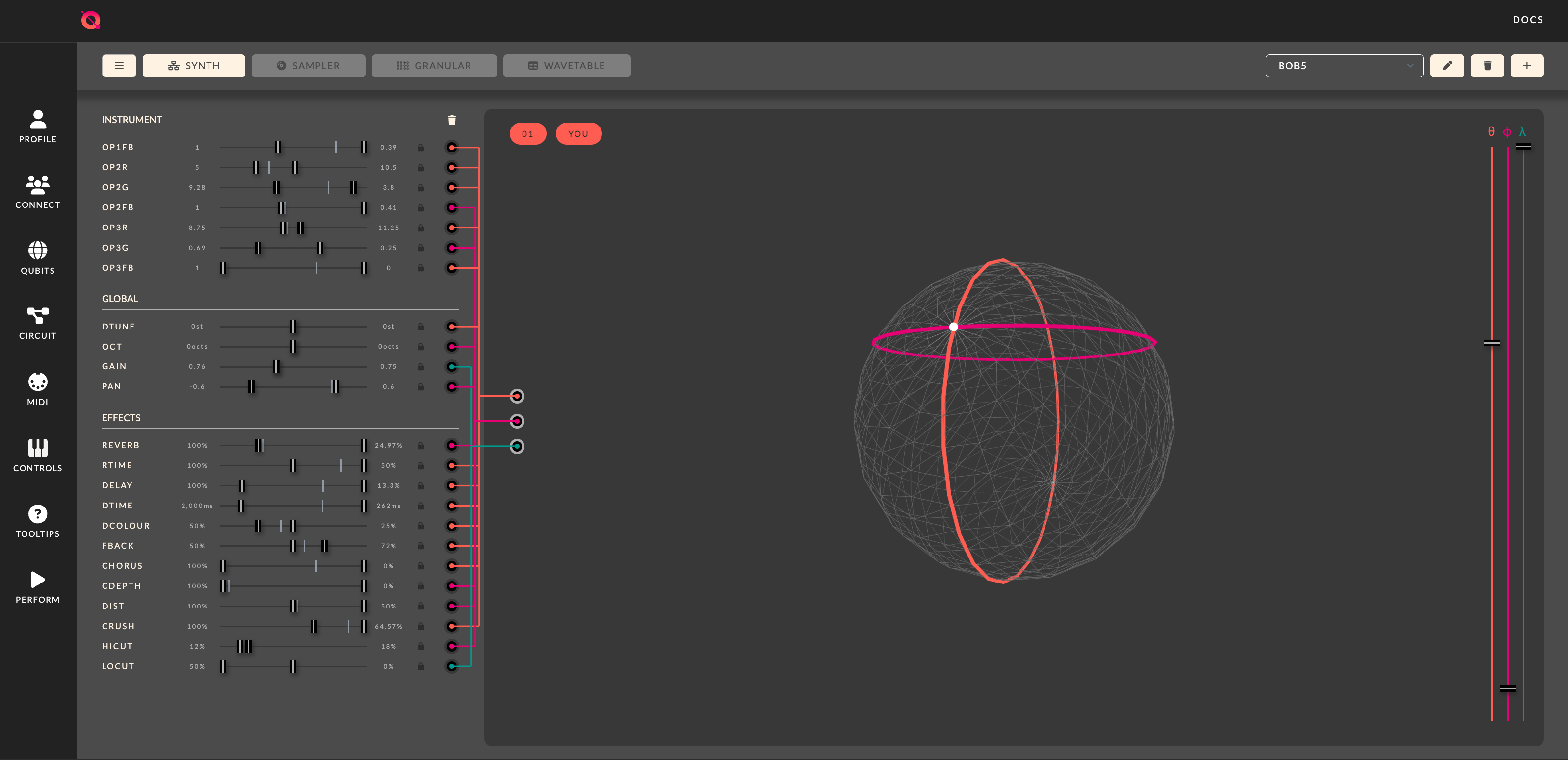}} 
   \caption{The qubit depicted within Actias. \bM Rotation and measurements of the--either physical or simulated--qubit are MIDI controlled.   For example, in quantum guitar \cite{quantumguitar}, this is done by means of foot pedals.  The state of the qubit is synthetically sonified in Actias.\e}  
    \label{ActiasQubit}
\end{figure}

As a running example `instrument' we will use one or multiple sonified qubit(s), which can be achieved through Moth's Actias synth, that was inspired by the earlier Q1Synth \cite{miranda2023q1synth}.  Actias' qubit is depicted in Fig.~\ref{ActiasQubit}. Actias can be downloaded here:
\begin{center}
 \href{https://github.com/moth-quantum/actias-electron}{https://github.com/moth-quantum/actias-electron} 
\end{center}
\begin{center}
 \href{https://github.com/moth-quantum/actias-backend}{https://github.com/moth-quantum/actias-backend} 
\end{center}
New versions are currently under development by other parties such as Quantum Entertainment Berlin, and will remain open-source.

\begin{figure}[h!]
   \centerline{\includegraphics[height=100pt]{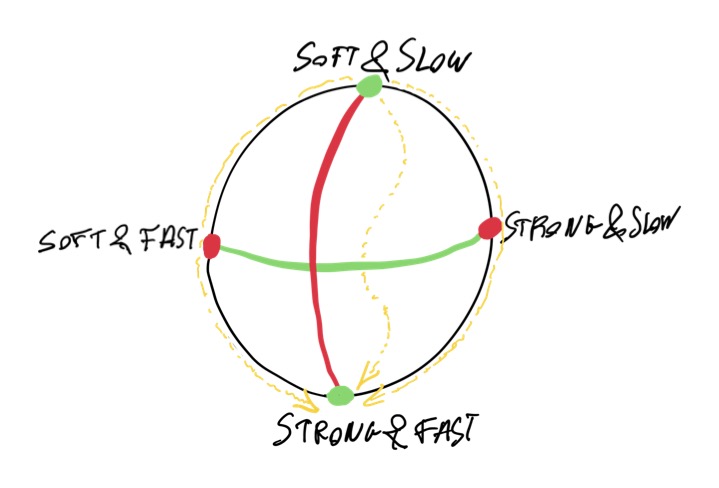}\vspace{-2mm}}
   \caption{Mentally created \bM `half'\e qubit for a Grand Piano.  There are many paths for going from (Soft and Slow) to (Strong and Fast) as indicated in yellow. \bM A binary variable could turn this into a full qubit, but half a qubit already posses all quantum essential properties.\e}
    \label{PianoQubit}
\end{figure} 

At the same time, all of the notation for a concrete qubit also applies to a mentally created qubit-like system, like the one that we describe in Section \ref{sec:pianoqubit},  and  is depicted in Fig.~\ref{PianoQubit}.  This qubit-like system can be performed  on a grand piano.\e

\begin{figure}[h]
   \centerline{\includegraphics[height=180pt]{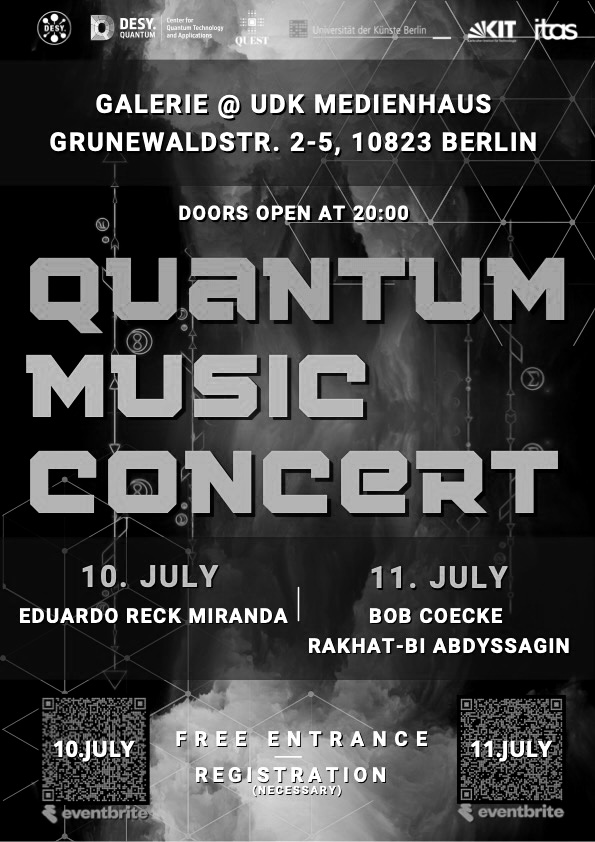}\includegraphics[height=180pt]{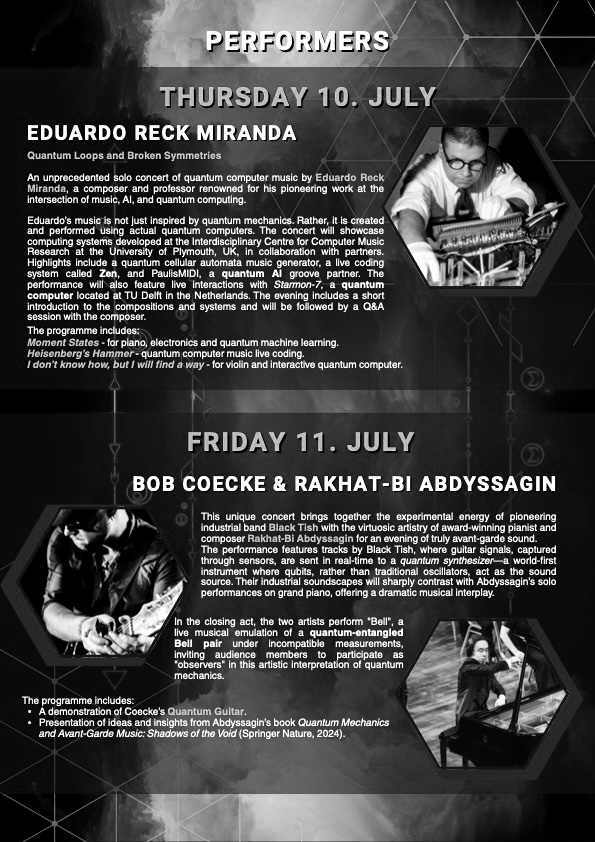}} 
   \caption{Poster from a performance in Berlin combining Quantum Guitar and Grand Piano.} 
    \label{fig:Berlin}
\end{figure} 

\bM The authors have already done  performances with these two kinds of qubits `mentally entangled', on two occasions, including the one in Fig.~\ref{fig:Berlin}. \e


\subsection{Earlier related work in quantum composition.} 

\begin{figure}
   \centerline{\includegraphics[width=200pt]{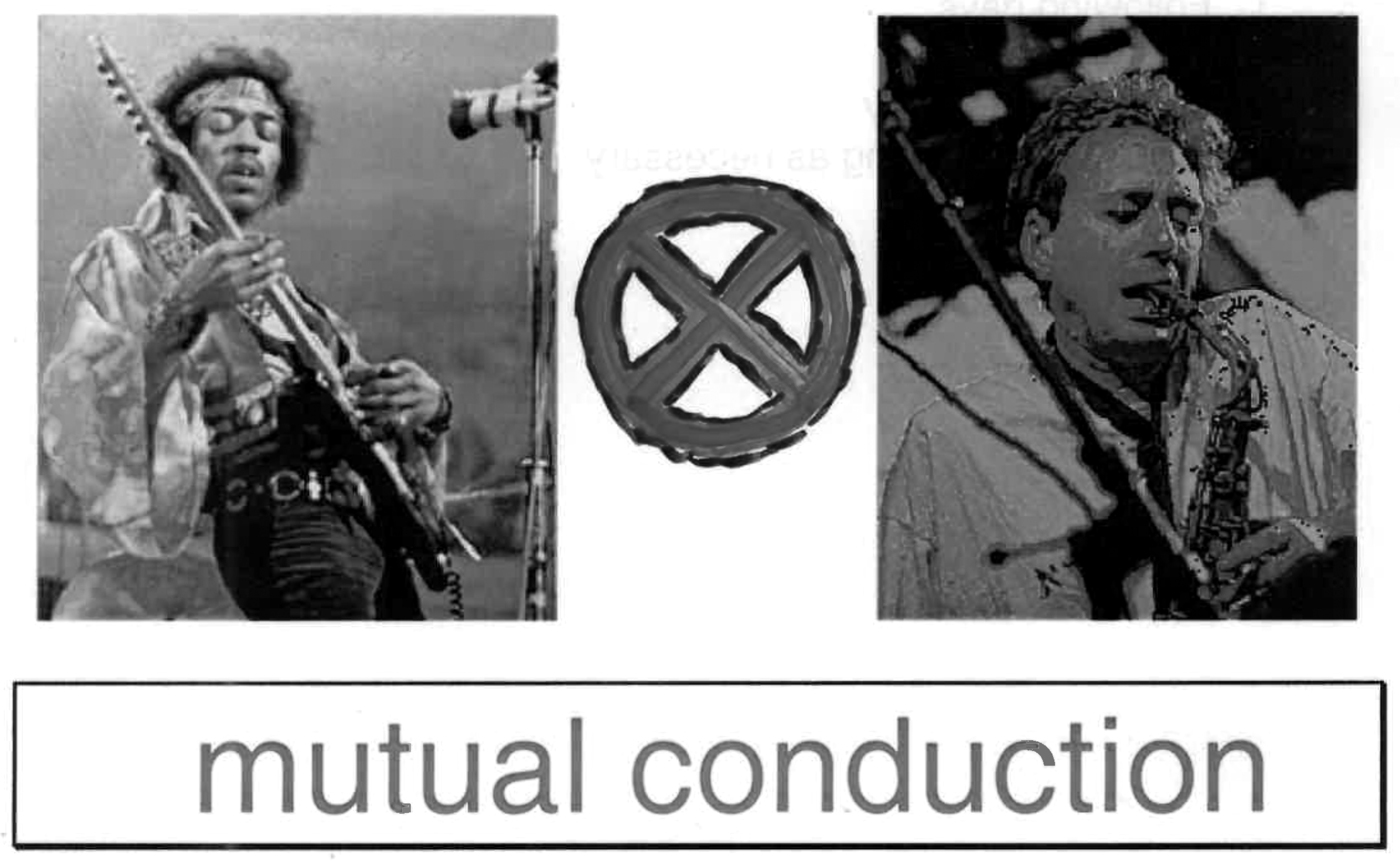}}
   \caption{Slide from a talk by Coecke at the School of the Art Institute of Chicago in the late 90's, suggesting the use of quantum interaction as a new way for musicians to interact with each other.  \bM As discussed in Section \ref{sec:quantumway}, the tensor symbol in this picture can be taken quite literally, when one imagines a monoidal category in which the objects are musicians, and the states constitute their respective musical output.  The ways in which they interact are encoded in the morphisms, possibly as described in this paper. 
   These ideas occurred to us during  collaborations with the late jazz musician/composer Eddy Loozen, in the late 90's.  
   In the picture are Jimi Hendrix and John Zorn.\e
   }
    \label{fig:Chicago} 
\end{figure}

The suggestion for bringing \bM interaction in\e music to the forefront in \bM in the form of quantum structure\e goes back to talks by one of the authors in the late 90's as seen in Fig.~\ref{fig:Chicago}, \bM which is a slide taken from a talk at the Art Institute of Chicago.  At the time, some small scale performances were done using implementations in IRCAM's MAX, but they were not recorded.  \e 

The first music generated \bM by quantum computation on a quantum computer\e, meanwhile some 5 years ago,  \bM called ``Bob's Cigar Buzz" \cite{miranda2021bobscigarbuzz}, also relied on  CQM/QPic.  \e  It was achieved by a team under the pseudonym Ludovico Quanthoven \cite{quanthoven}, and as the title of the paper states, the approach was based on Quantum Natural Language Processing (QNLP) \cite{QNLP-foundations,QNLPPlus100}, which in turns, is \bM a canonical application area\e of CQM/QPic  \cite{CSC,teleling,qspeak}. More specifically, the pictorial linguistic structure of QNLP was replaced by musical structure, and the linguistic meanings were replaced by musical meanings.  The QNLP software package {\tt lambeq}  \cite{kartsaklis2021lambeq} was in that way  turned into a quantum music software package {\tt Quanthoven}.  So from very early on quantum music has been directly connected to CQM/QPic.  

\bM ``Zeno" composed by Miranda in 2018 \cite{miranda2018zeno} used the output of a clarinet on a quantum computer to accompany the clarinet.   So the quantum computer acted as a buddy musician who improvised with the clarinet tunes, as if in a jazz-like gig. \e

There are many works in quantum music that also used the concept of entanglement, as a formal apparatus \cite{putz2015quantum,putz2022perception}, as a physical resource \cite{dediosrodriguez2025sound},  as a instrument-design principle \cite{carney2025qubit,carney2023quid}, and in the HCI/post-humanist sense \cite{mice2023chaos,waters2021entanglements,frauenberger2019entanglement}.  In \cite{mannone2021gestart} the authors also use category theory in quantum music.

\begin{figure}
   \centerline{\includegraphics[width=180pt]{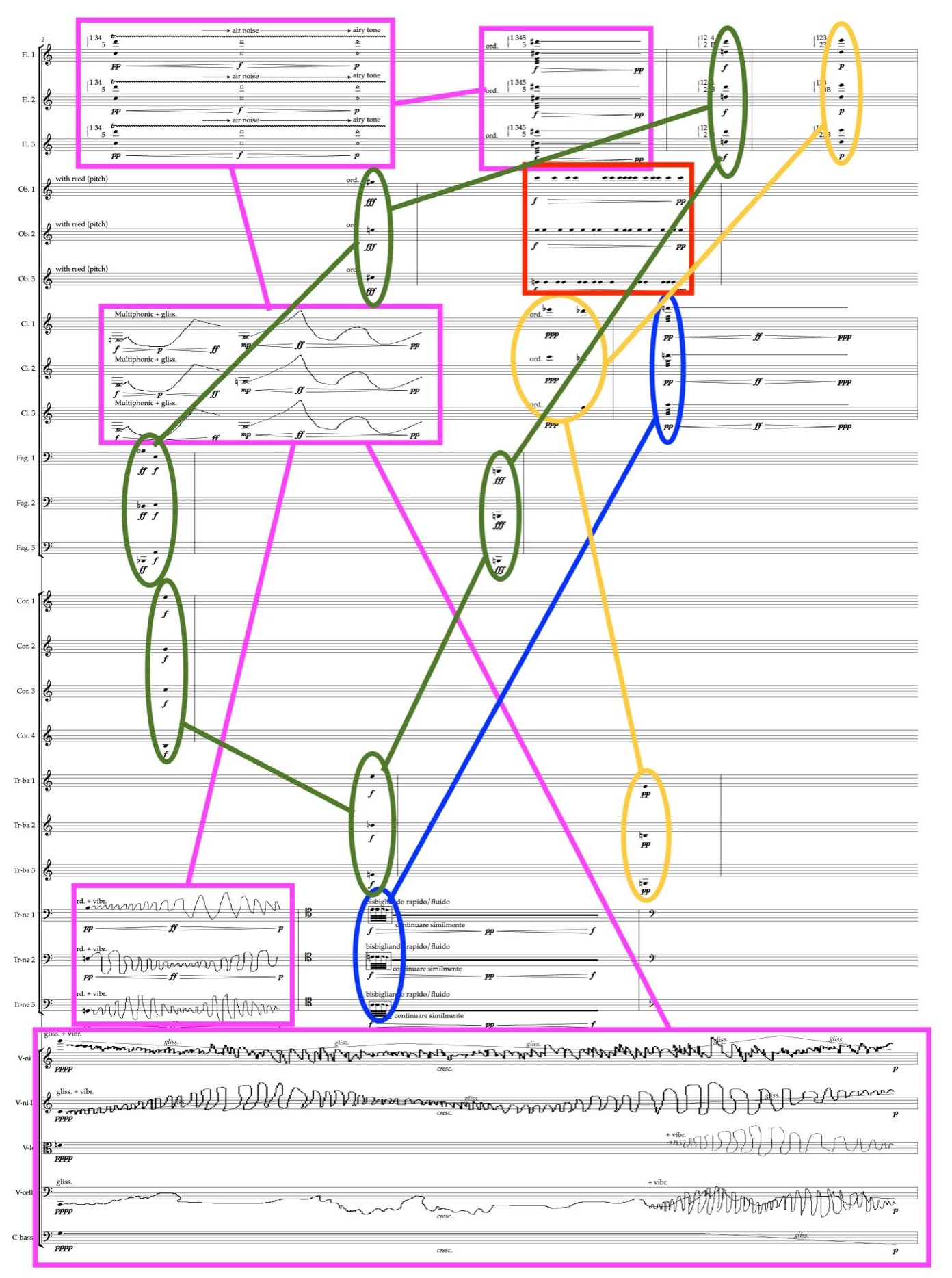}}
   \caption{A page from the score of Abdyssagin's Symphony No.~2 ``Chaos and Order" (2023), indicating the relational nature of music, in a way that instantiates part of the formalism introduced in this paper \cite{AbdyssaginChaosAndOrder,AbdyssaginPhDStAnd}. \bM Here the diagrammatic tools are used as a means of structural analysis, in particular by highlighting in different colours and shapes the relations and interconnections as well as similarities and equivalences between sonoristic classes and timbre-texture sets. More on this (together with the conceptual and technical explanation of details) can be found in the recent book \cite{AbdyssaginLogic}. This analysis doesn't involve the quantum features that we introduce in this paper. \e}
   \label{fig:Rakhat} 
\end{figure}

The relational nature of music is also clearly visible in the structural diagrammatic analysis of the score of Abdyssagin's Symphony No.~2 ``Chaos and Order" (2023) \cite{AbdyssaginChaosAndOrder,AbdyssaginPhDStAnd} in Fig.~\ref{fig:Rakhat}. \bM Here the diagrammatic tools are used as a means of structural analysis, in particular by highlighting in different colours and shapes the relations and interconnections as well as similarities and equivalences between sonoristic classes and timbre-texture sets. \e

Examples of how quantum phenomena are reflected in music at the level of composition and numerous other correlations can be found in Abdyssagin's book 
\cite{AbdyssaginQMandAM}. \bM Abdyssagin's recent ``Quantum Universe" Symphony for Organ, Quantum Guitar and Electronics (2025) exemplifies these. It is the first quantum-theory based musical composition.   Moreover,  for the first time one of the oldest instruments in the world, pipe organ, got compositionally `entangled' alongside one of the newest instruments, a quantum guitar connected to a quantum computer, where operations on qubits transform sound in real time. In that sense it is directly related to this paper, as an overall quantum conceptual framework comes together with quantum instruments.
It is also an example in music where theory has been implemented in the form of an experiment as a musical composition. It opened the Oxford Quantum Centenary Conference on 20th November 2025 in Merton College Chapel, Oxford \cite{QU}. \e  

\bM 
\section{Quantum(-like) in many ways: formalism and implementation}\label{sec:quantumway}

The work introduced in this paper is quantum or quantum-like in many ways.

Firstly, we  make use of an incarnation of CQM/QPic, namely the ZX-calculus \cite{CD1,CD2}, developed by Coecke and Duncan in 2007, and now prominent in quantum foundations and quantum technology.  It is a substitute for the Hilbert space formalism of quantum theory, as it has been proven to be a universal \cite{ContPhys} and complete \cite{hadzihasanovic2018two} graphical language for quantum system.  This means, everything that is expressible in Hilbert space quantum theory can be expressed in the ZX-calculus, and any equation derivable in Hilbert space, is derivable in the ZX-calculus. So we effectively are using the quantum formalism,  and in Section \ref{sec:filling} we show how our notation `fills' the full tensor product space.  We can now also use the rules of ZX-calculus to compute musical states.

That said, as a category-theoretic language, and more specifically, monoidal category-theoretic language \cite{AbrLNP,BaezLNP,CatsII}, all one needs to specify are processes and parallel and sequential composition thereof. As such, monoidal categories accommodate many kinds of process-theoretic phenomena \cite{CKbook}, for example, in \cite{ContPhys,CatsII} an example with food and cooking is provided.  A naive monoidal category of musical score could consists of snippets of scores for individual instruments, their sequential composition, and the parallel composition of instruments.  Alternatively, it could have performers as objects, with their musical outputs as states, their progressions as morphisms, and their interactions provided by the tensor structure--see Fig.~\ref{fig:Chicago}.

For the specific case of ZX-calculus, essential quantum features, such as complementary spiders and the phase-group, should be accounted for within the musical interpretation. The notion of quantum measurement then automatically arises, given that measurements are also spiders. The inherent quantum non-determinism is also implicit within the spiders.  In Section \ref{sec:quantum} we introduce quantum measurements within our notation.  A standard reference for all of this is \cite{CKbook}.  A `mathematics-free' presentation is in \cite{QiP}.

Without going into technical details, new models can arise by varying the phase groups within the ZX-calculus, and in this manner we can find many other interesting musical models that still have essential quantum features. An analogy here   is passing from an ordinary qubit to Spekkens' toy qubit \cite{CES,MiriamSpek}.

Concretely, and this is a 2nd level of quantum-nativeness, we provide musical realisations of qubits, in Section \ref{sec:Bell1}.  Quantum Guitar \cite{quantumguitar} can operate with actual physical qubits, through the Actias software.  As explained in \cite{quantumguitar} and here in Section \ref{sec:qgit}, Quantum Guitar is a quantization--in the sense of Deutsch--of a guitar fretboard \cite{Deutsch2004,Pienaar2011,Singh2019}.  The second example is a mental representation of quantum complementarity on a grand piano, that we describe in Section \ref{sec:pianoqubit}.

In principle, when taking two quantum guitars, even the entanglement could be physical, although this has not yet been implemented in the Actias software.  We are planning to do so, as this is a fairly simple extension. Doing so, the entanglement could even arise from actual physical qubits.

An important point to re-iterate is that while our notation at first may look like classical constraints, it has essential quantum features built-in.  The wires represent entire Hilbert spaces of mixed states, so that the non-determinism resulting from measurements flows through the entire composition. Even at a more basic level, actual changes of musical state are imposed by  gates--See Section \ref{sec:filling}. In that sense, there is a fundamental dynamical component within the notation, distinguishing it from the structural analysis like in in Fig.~\ref{fig:Rakhat}.

\e

\bM \section{Relations in scores}\label{sec:basics}\e

\begin{figure}
   \centerline{\includegraphics[width=165pt]{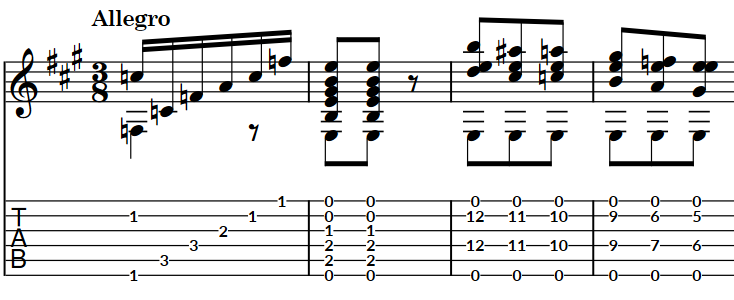}}
    \caption{Traditional music notation and corresponding guitar tablature, which specifies where to play particular notes--there may be up to 6 different ways to do so, affecting  playability and sound texture.} 
    \label{tabs}
\end{figure}

The notation introduced below augments a variety of existing music notations, 
including traditional staff-notation, guitar tablature, piano-roll (cf.~Figs.~\ref{QU} and  \ref{tabs}), 
as well as other modern forms of music notations. In fact, it is not even restricted to music notation, and could likely also apply to notations for dance, animated art, games etc. Here we will mostly introduce everything relative to traditional staff-notation.

We will mainly focus on the relationships between different instruments. For notes, as spider dots, being connected by a \bM plain\e wire expresses that these notes are the same. Hence, at this point the notation doesn't add any content, \bM as it merely makes something explicit that is already implicit the case\e. 
\beq\label{staff1}
\raisebox{-1.3cm}{\epsfig{figure=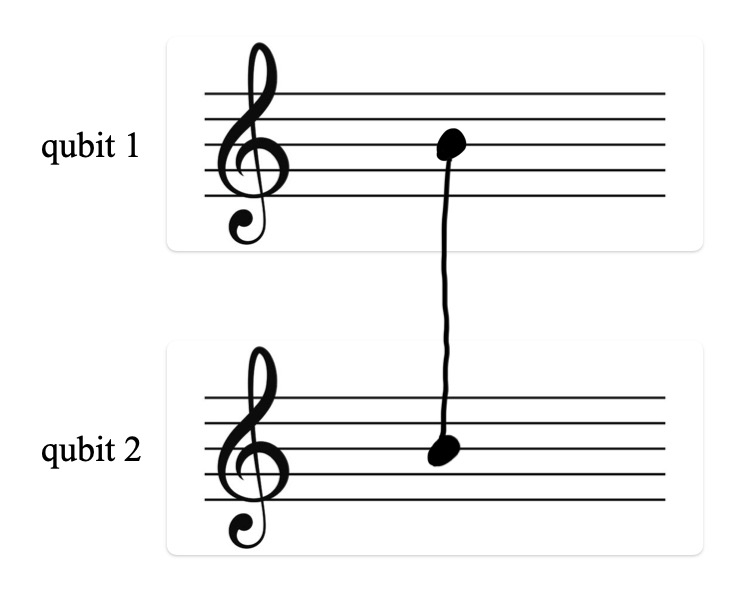,width=100pt}}
\eeq

A bigger `blob' around the staff means that whatever is in the blob is the same. This can either be specified, or a variable, e.g.,~for improvisation, or it can be a high-level abstract structure that can be further refined.
\[
\epsfig{figure=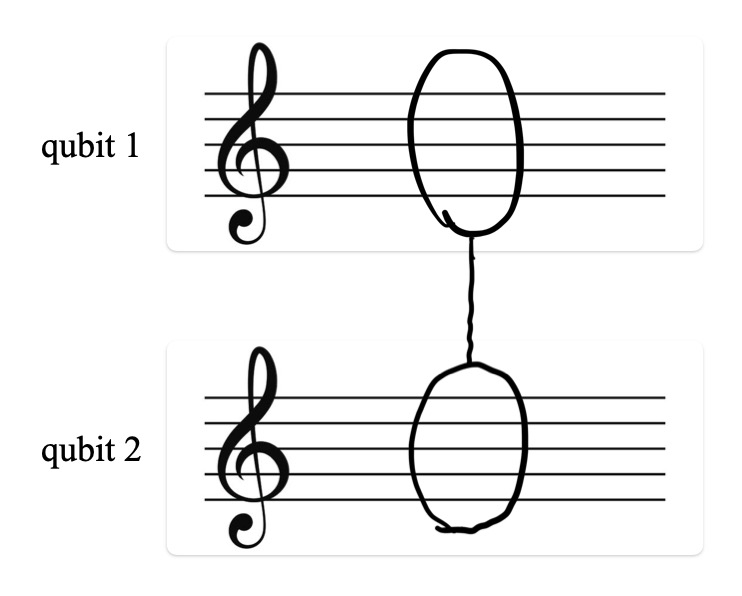,width=100pt}
\]

So such a plain spider gate means doing the same, as in an identical relation between what is in the circles. A gate on the wire allows one to express alternative relations between what is in the circles, of which there could be many variations, for example, a $\sharp^n$-symbol may indicate `$n$  half-tones higher'.
\[
\epsfig{figure=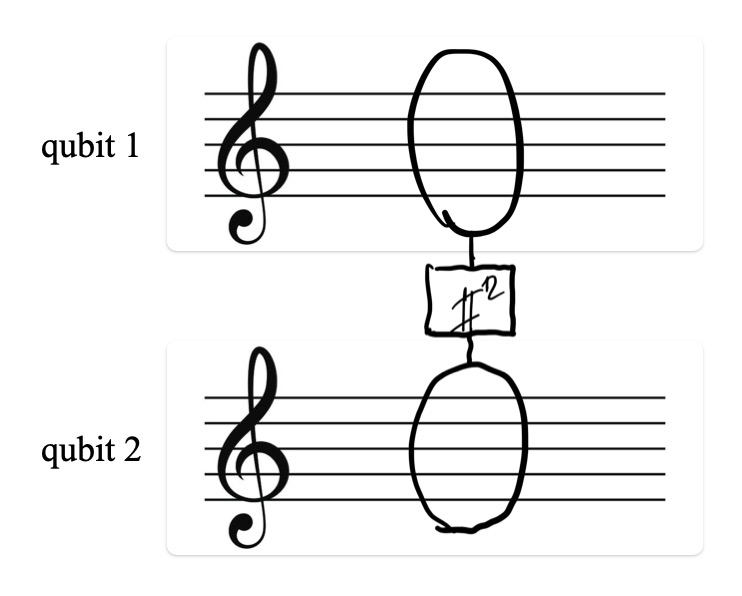,width=100pt}  
\]
There may be other relations that are defined in terms of a ``meta-glossary", for example indicating what one means by ``opposite/dual" etc. The gate can also be left as a variable, which in a live performance could be incarnated by a third party. The gate can also be taken to be a genuine quantum gate, that indicates relationships such as complementarity, by means of a Hadamard gate, \bM as we discuss in Section \ref{sec:filling}. \e

\bigskip

\begin{remark}
Thus far we took ``sameness" to be identical, but just like in mathematics (cf.~same vs equivalence vs isomorphic), there can be many notions of sameness, for example, ``alike", ``inspired by", ``anticipates", ``follows" etc. Specifying this could be part of the meta-glossary. Possibly, if different forms of sameness are required in one score, this could also be indicated, e.g.,~using ``$\simeq$" notation.
\[
\epsfig{figure=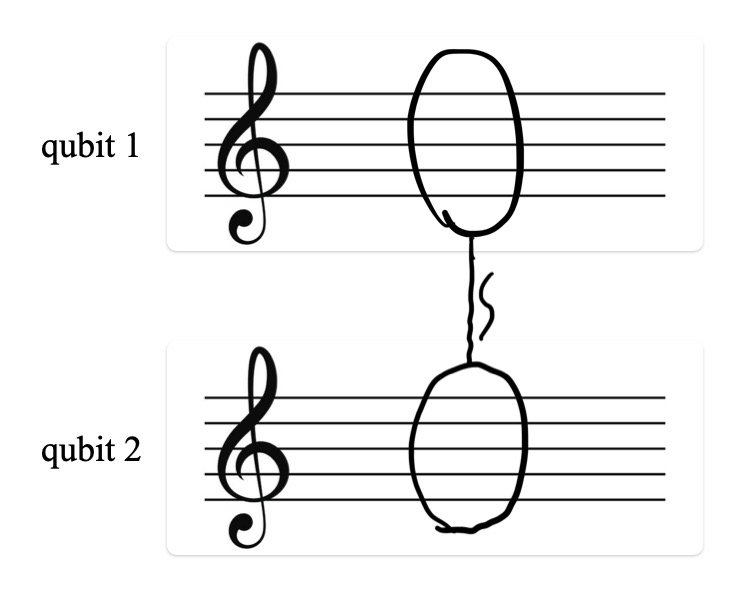,width=100pt}   
\]
One may also want to consider partial notions of sameness, e.g.,~sameness only at the level of notes, or sounds, or rhythm, which could be indicated by extra notation in the gate.  Alternatively, this could again be subsumed in the meta-glossary. A meta-glossary will serve the function similar to the role of performance notes/legend in contemporary music scores where many extended instrumental techniques need to be described and elucidated. 
\end{remark}

\section{Lead and follower in scores}\label{sec:MS}

In improvised live performance, sameness could mean that one musician takes the lead and the other one follows. And this, of course, can change during the performance.  This can be indicated by a tilted gate.
\[
\epsfig{figure=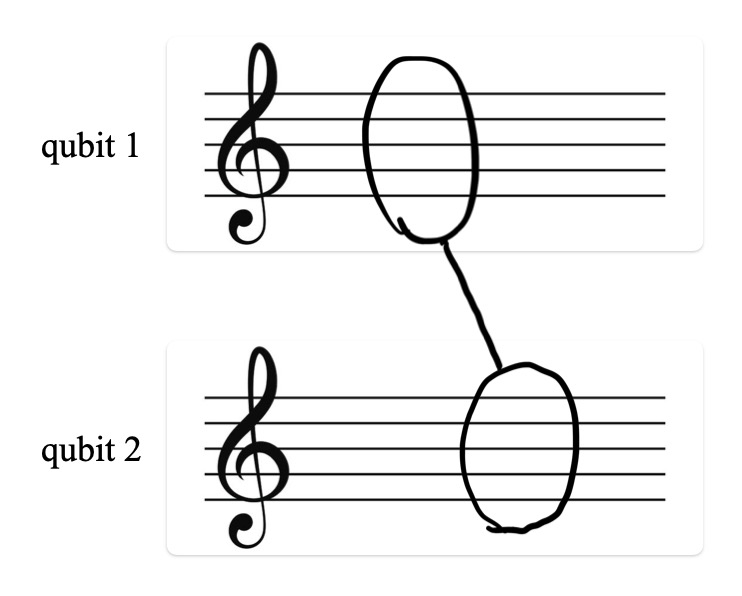,width=100pt}   
\]
Alternative indications could be an arrow. The above however reflects the cause-consequence relation in a way that is more aligned with physics by providing a temporal connotation. 

\bM \subsection{Logical vs.~operational reading}\e

In ZX-notation, in principle, all gates are always tilted given that each wire is assigned to either be an input or an output. However, since the two possible tilts turn out to be equal by the axioms of the calculus \cite[Lemma 9.46]{CKbook}, one denotes it as non-tilted in order to stipulated that equality.
\beq\label{eq:tilt1}
\raisebox{-0.65cm}{\epsfig{figure=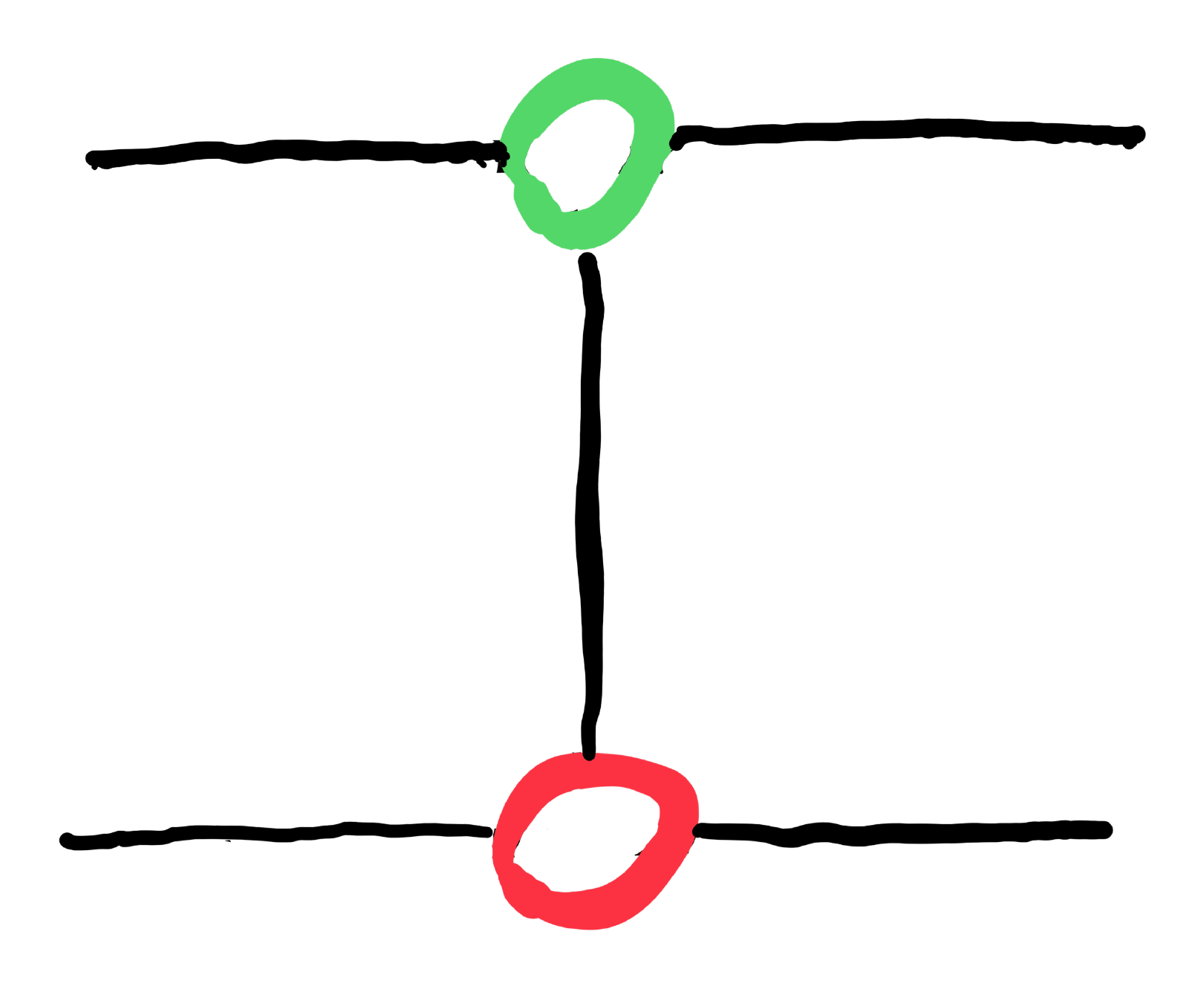,height=40pt}}\ \ := \raisebox{-0.65cm}{\epsfig{figure=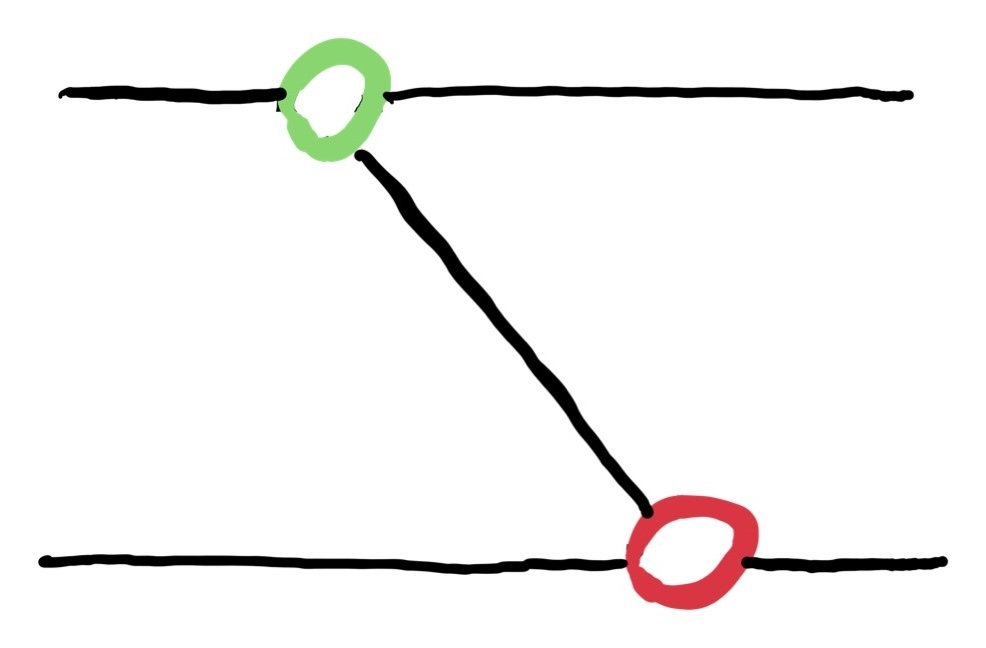,height=40pt}}\ \ = \raisebox{-0.65cm}{\epsfig{figure=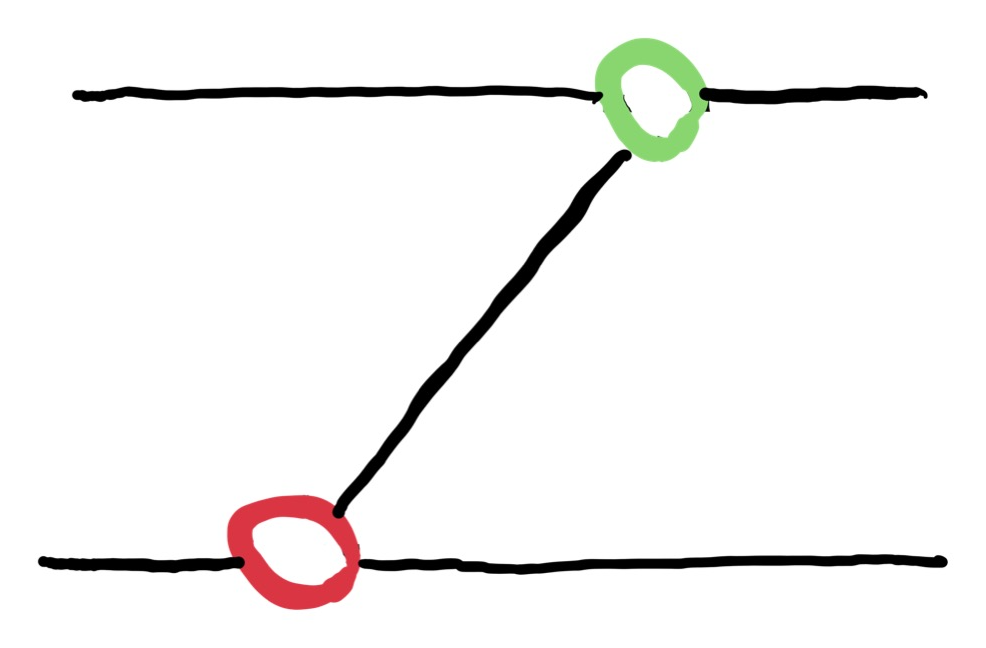,height=40pt}}
\eeq
\bM Conversely, the notation enables on to introduce the tilting, as a syntactic means that enables one to account for a relative temporal connotation.  So we make use of the dual reading of ZX-diagrams, the `logical' one, which is the one in which equational reasoning is done, and the `operational' one, in which operational implementation is specified--see \cite[\S 4.4.3]{CKbook}.  In eq.~(\ref{eq:tilt1}) we perform equational reasoning, while here in diagram (\ref{eq:tilt2}) we indicated operational execution:
\beq\label{eq:tilt2}
\raisebox{-1.20cm}{\epsfig{figure=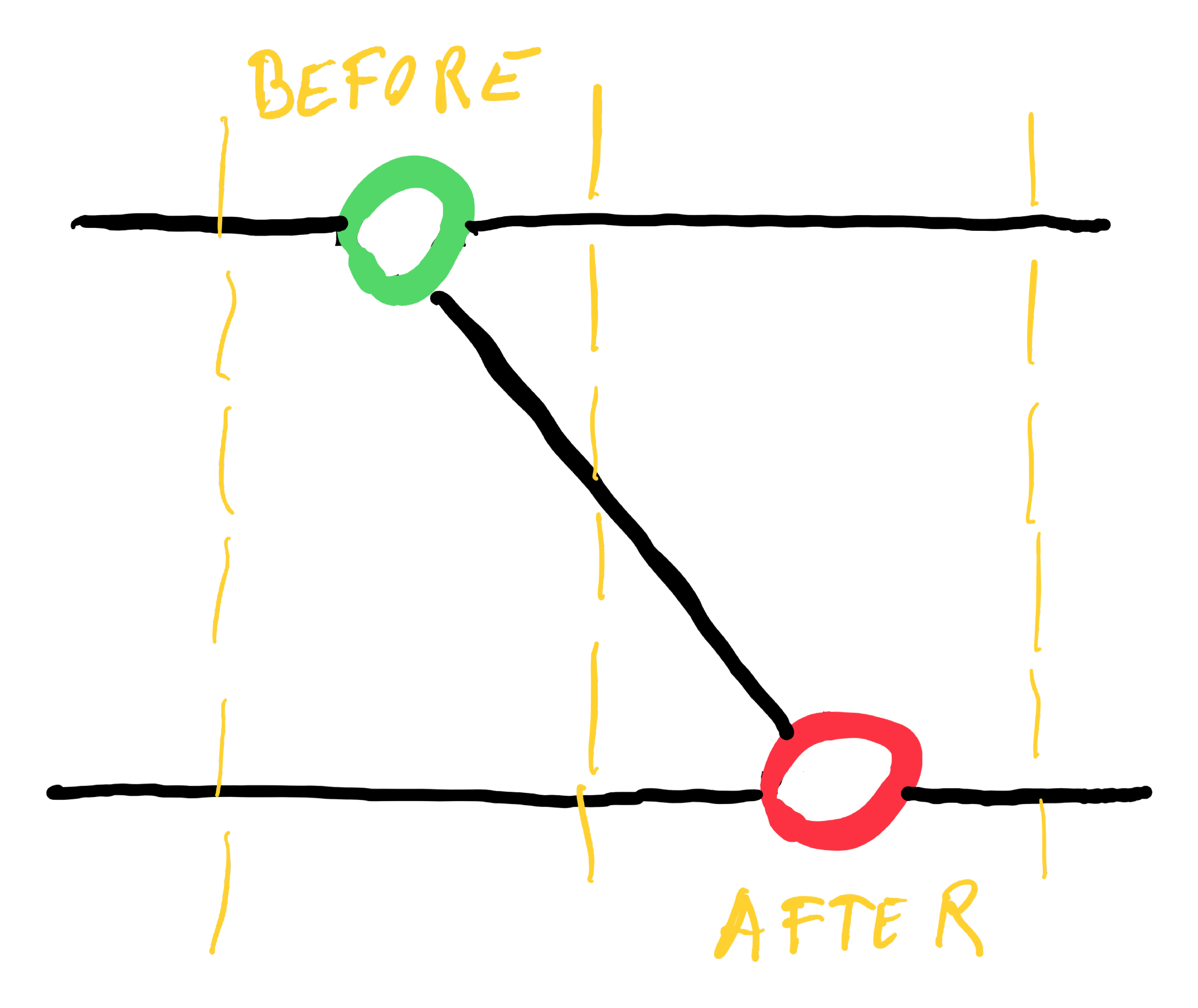,height=80pt}}
\eeq
As a convention, below, when equations are involved  the diagrams should be conceived logically, while if not, they should be read operationally.
\e

\bM \section{Quantum-native notation in scores}\label{sec:filling}

Following on on the above, we can introduce quantum-native gates, that will result in `filling' the entire tensor product space.  Assuming that the musical state is described by a qubit, like in Actias,  we can use any qubit gate, for example, a phase gate:
\beq\label{eq:filling}
\raisebox{-1.3cm}{\epsfig{figure=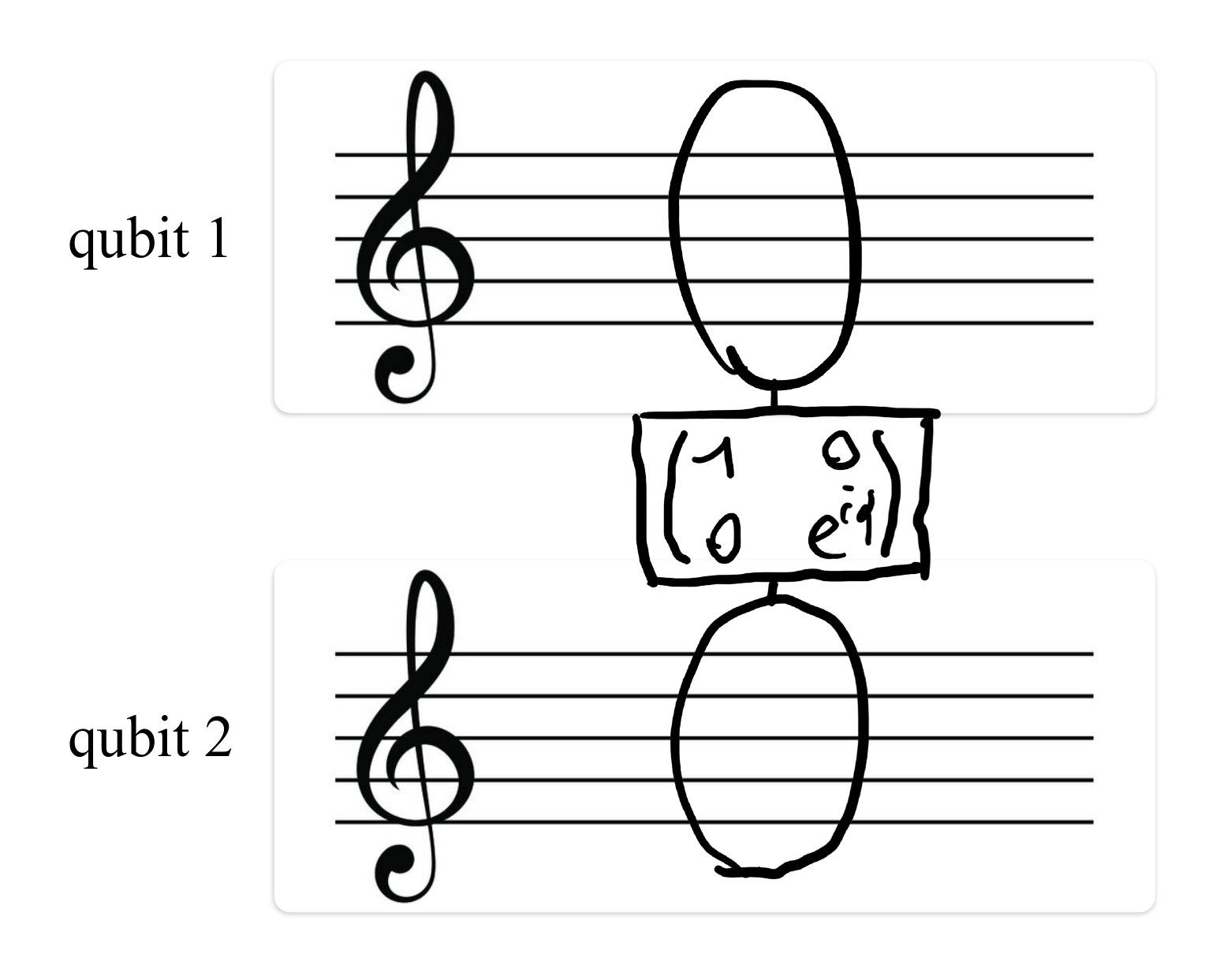,width=100pt}}    
\eeq
In this way, we can obtain any entangled state as  musical relatedness.

\begin{lemma}
If the musical states are described by qubits, diagrams like the one above enable one to describe any two-qubit states.   Using quantum circuits, any maximally entangled two-qubit state could be realized.
\end{lemma}

\begin{proof}
This follows directly from the isomorphism between (a) linear maps between two qubits, and (b) the entangled states of two qubits. The 2nd statement follows by the fact that unitary gates in score \ref{eq:filling} generate maximally entangled states, and that all such gates can be realized by a unitary quantum circuit.
\end{proof}

As a consequence, the notation encompasses the interaction of two sonified qubits through quantum entanglement.  As an example in Dirac notation, taking the one-qubit gate in score (\ref{eq:filling}) to be a Hadamard gate, then the two-qubit gate is the CZ-gate.  Letting the initial states be $|+\rangle$ and $|0\rangle$, we obtain an entangled state:
\[
CZ  |+0\rangle = |0+\rangle+|1-\rangle
\]

This example also clearly shows that the notation is not just a constraint like in Fig.~\ref{fig:Rakhat}, but an operation that alters the music. In order to compute a musical state at a certain instance of the score, we will now be able to use the rules of ZX-calculus.  

While here we used strict Hilbert space notation, all of the above can also be adopted to a mentally created qubit like we describe in Section \ref{sec:pianoqubit} and is depicted in Fig.~\ref{PianoQubit}.  We provide an example of this in Section \ref{sniped}.
\e

\bM \section{Complementary  measurements in scores}\label{sec:quantum}\e

Now we introduce the notion of quantum measurement into music notation, for the purpose of composition and/or performance.  This means that we will have to distinguish between `thick' or `doubled' quantum wires and blobs, vs `thin' or `single' classical wires and blobs, as is done in CQM/QPict  \cite{CKbook,CQMII}.  Specifically, measurements constitute the passage from thick quantum wires to thin  classical wires, and doing the opposite represents encoding of classical data as quantum data \cite{CKbook,CQMII}. For all that we said above, the wires of staff and gates could all have been the same, either quantum or classical, so we ignored the issue. Below we will explain how measurements and quantum gates can be used together in a score without any ambiguity, which is for example necessary when one wants to entangle the qubits before measuring.  

For the purpose of composition and/or performance, a measurement means the following, which here we take to be on one of two qubits, that may be entangled:
\bit
\item A \underline{choice} of a measurement is made for one of the qubits, which includes \underline{specification} of what the two eigenstates of the measurement are, for example, a $Z$-measurement with eigenstates $|0\rangle$ and $|1\rangle$.  This is done by whatever/whomever plays the role of the observer.
\item They then \underline{decide} when to perform the measurement, say on qubit 1.
\item When the measurement is \underline{initiated}, the state of qubit 1 \underline{collapses} to the eigenstate corresponding to the measurement outcome.  This is a \underline{non-deterministic process}, but for which some mechanism may be provided. 
\item Depending on the relationship between qubit 1 and qubit 2, the latter \underline{collapses} to a corresponding state. In the case that the relation is ``being the same", this is the same state as qubit 1 collapses too.   
\eit
We can depict the measurement as follows:  
\beq\label{staff6}
\raisebox{-1cm}{\epsfig{figure=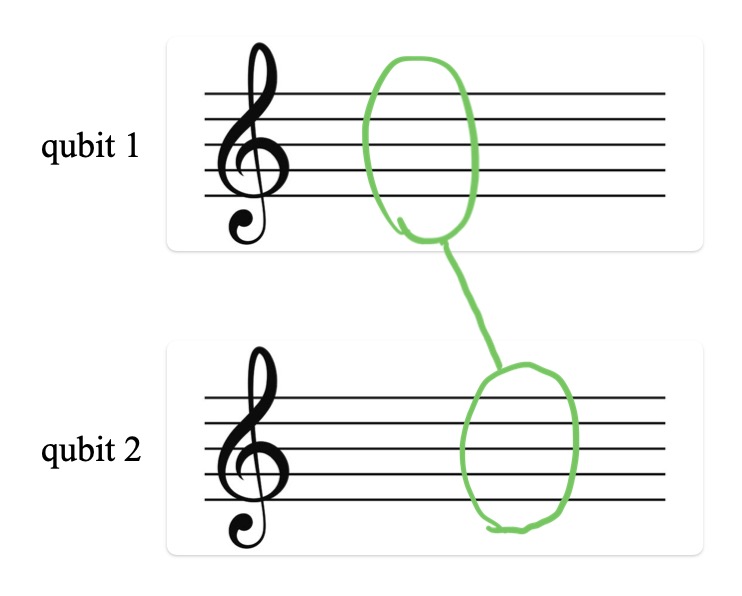,width=100pt}}   
\eeq
where the colour represents the choice of measurement.  A different measurement, say an $X$-measurement relative to a $Z$-measurement, could then be depicted as follows:
\[
\epsfig{figure=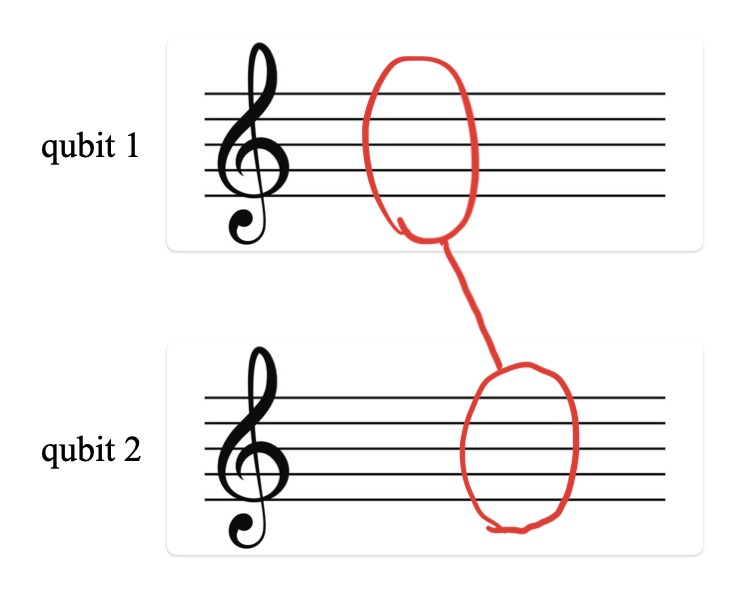,width=100pt}   
\]
The wire connecting the blobs indicates the influence of qubit 1 on qubit 2.  Strictly speaking, this implicitly assumes that the two qubits are entangled: 
\[
\raisebox{-1.28cm}{\epsfig{figure=staff6,width=100pt}}   \ \ := \ \ \raisebox{-1.28cm}{\epsfig{figure=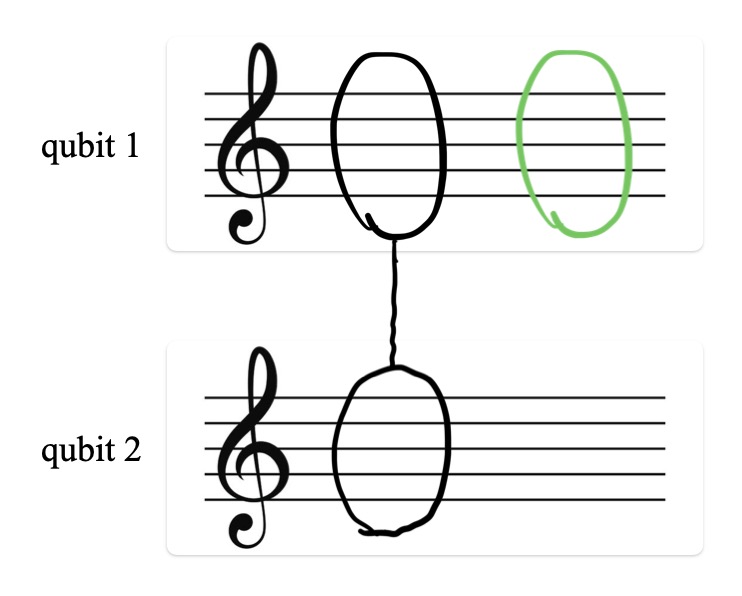,width=100pt}}   
\]
\bM where the black gate represents the entanglement, and the green blob the measurement\e. 
However, the more compact notation as in (\ref{staff6}) may be more appropriate for musical performance, indicating direct causal connections at the time they should be accounted for.  Translated to  the QPic formalism, the above equation becomes, when for simplicity instantiating to a specific note as measurement outcome:
\beq\label{eq:meas1}
\raisebox{-1.1cm}{\epsfig{figure=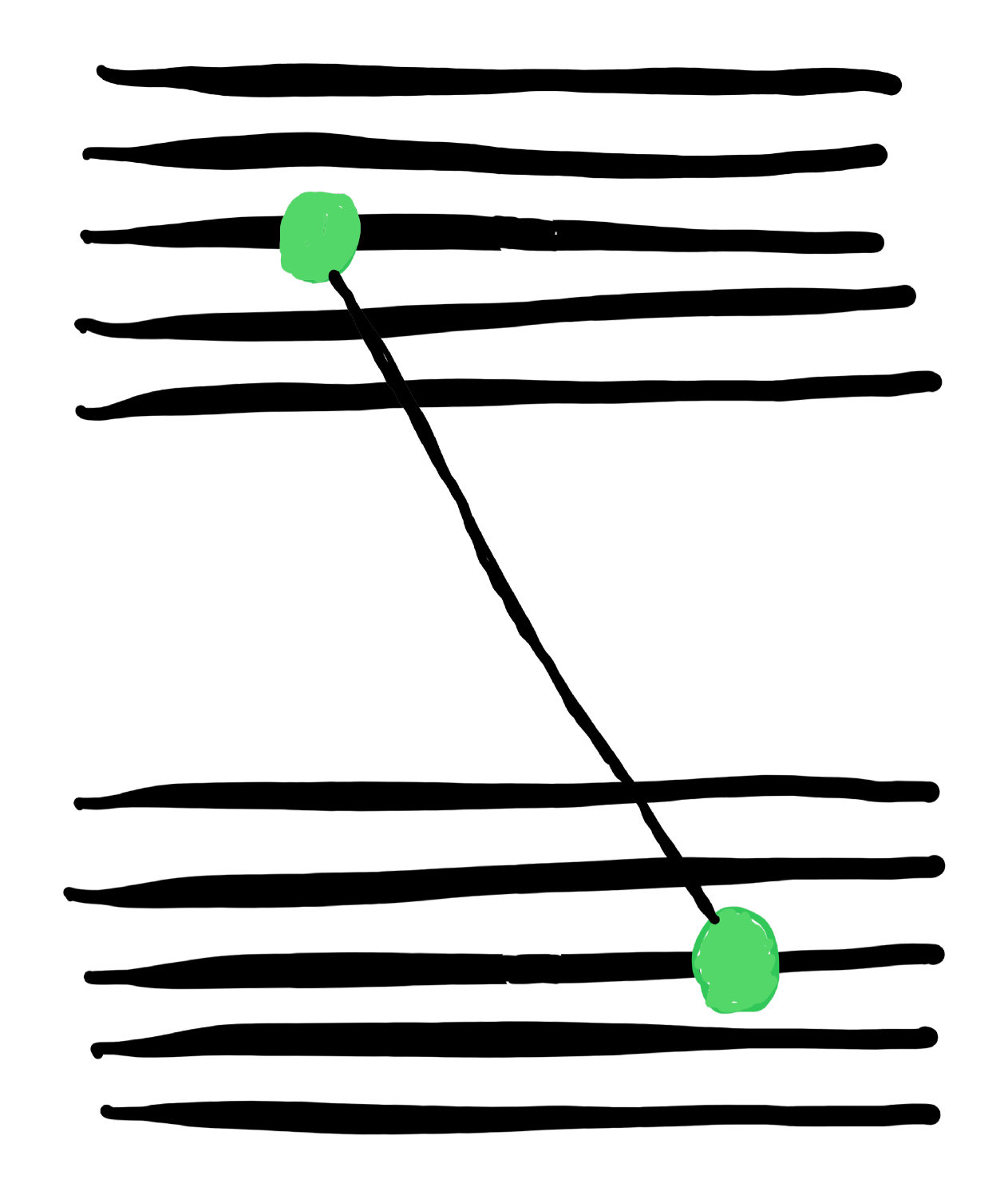,width=60pt}}   \ \ = \ \ \raisebox{-1.1cm}{\epsfig{figure=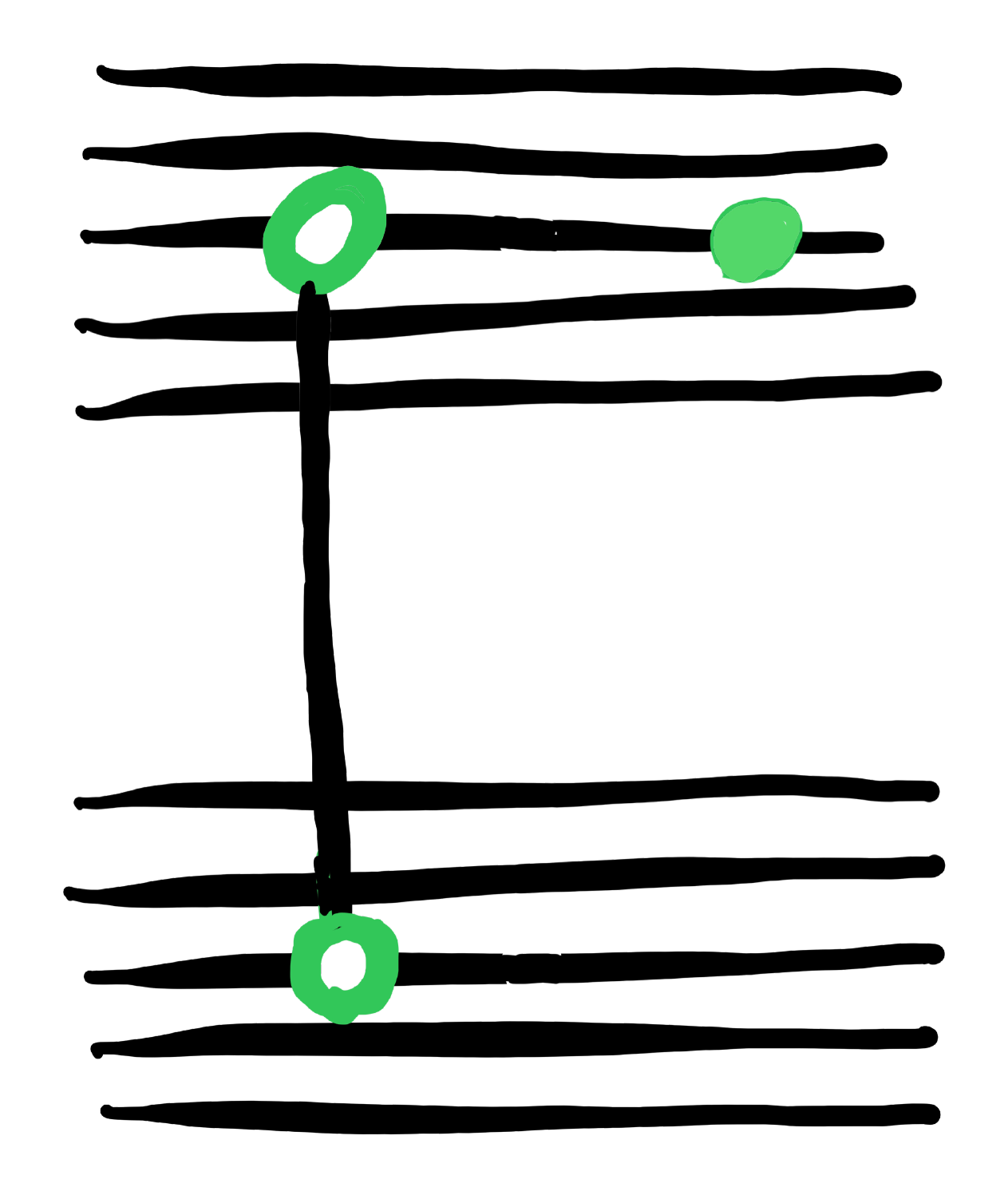,width=60pt}}   
\eeq
So this is where the thick vs.~thin wires and nodes becomes crucial. We now formally prove our claim.
 
\begin{lemma}
Equation (\ref{eq:meas1}) indeed holds.  
\end{lemma}

\begin{proof}
Using `bastard spider fusion' \cite[Section 8.3.3]{CKbook} we obtain: \vspace{-2mm}
\[
\raisebox{-1.1cm}{\epsfig{figure=ZXmeas1b,width=60pt}}   \ \ = \ \ \raisebox{-1.1cm}{\epsfig{figure=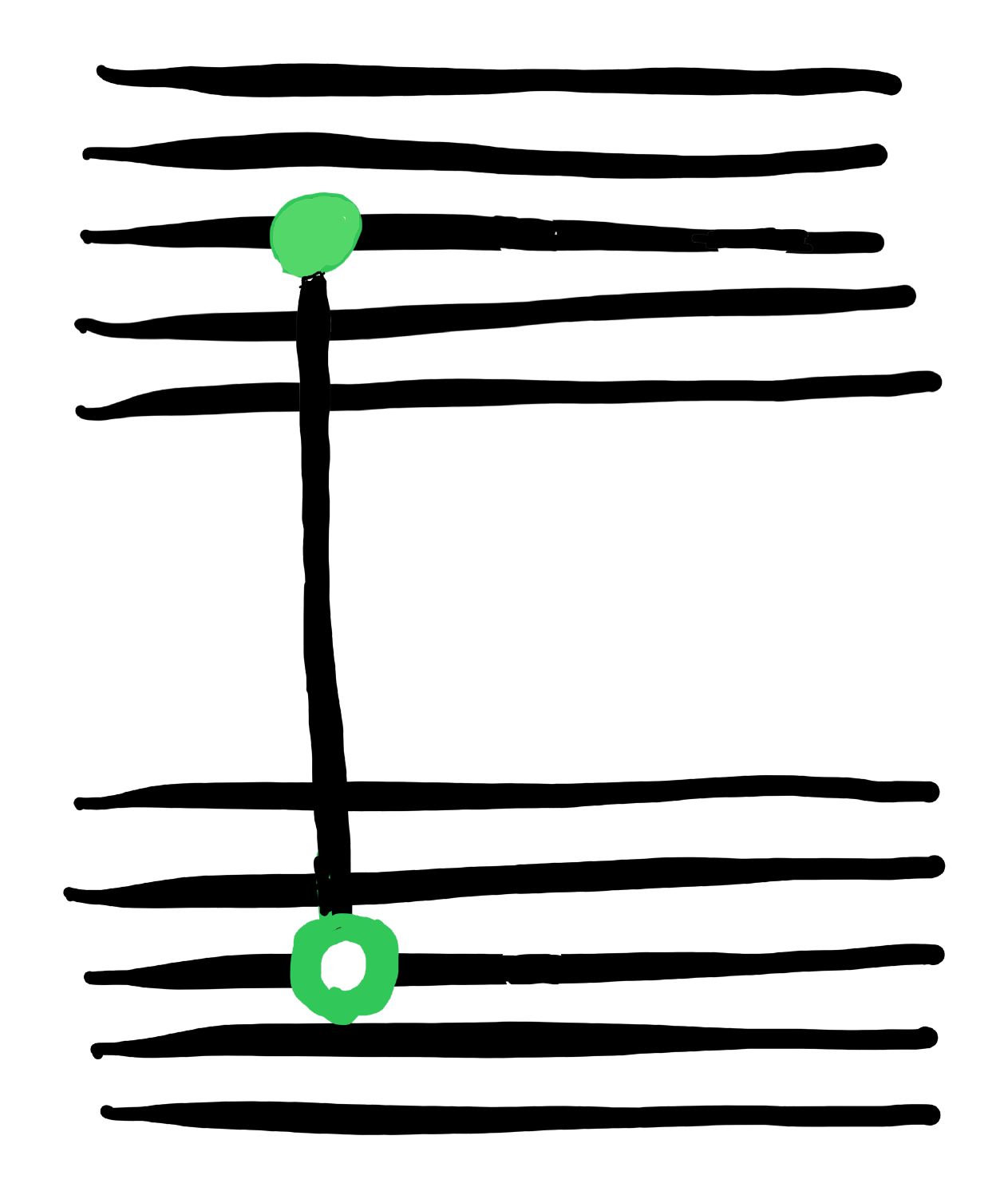,width=60pt}}      \ \ = \ \ \raisebox{-1.1cm}{\epsfig{figure=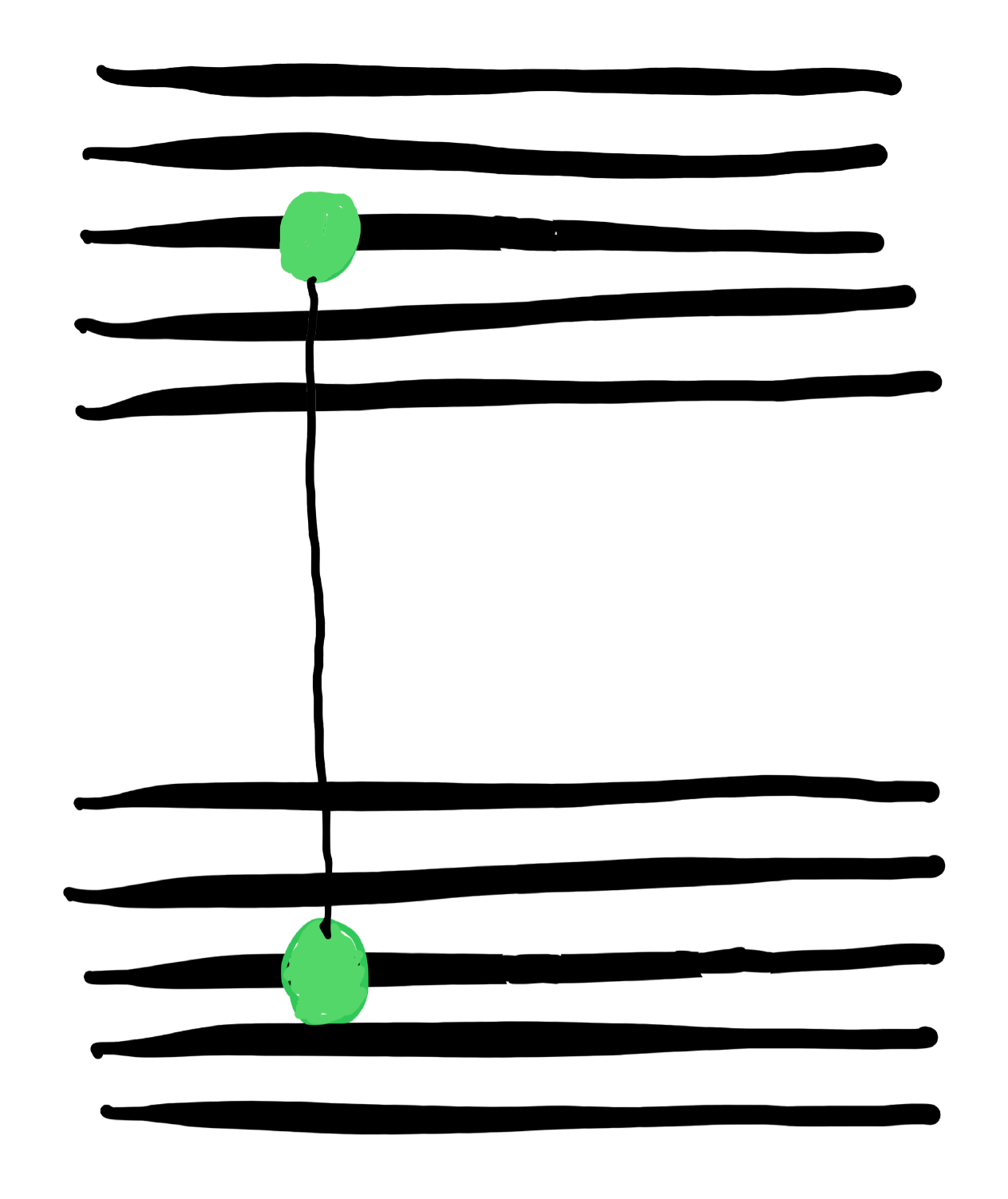,width=60pt}}      \ \ = \ \ \raisebox{-1.1cm}{\epsfig{figure=ZXmeas2b,width=60pt}}
\]
\end{proof}

\bM Putting everything above together, we can now also use the following notation, which involves measurements on qubits after having performed an entangling operation that alters the manner in which the measurement on the first qubit will affect the second\e:
\[
\epsfig{figure=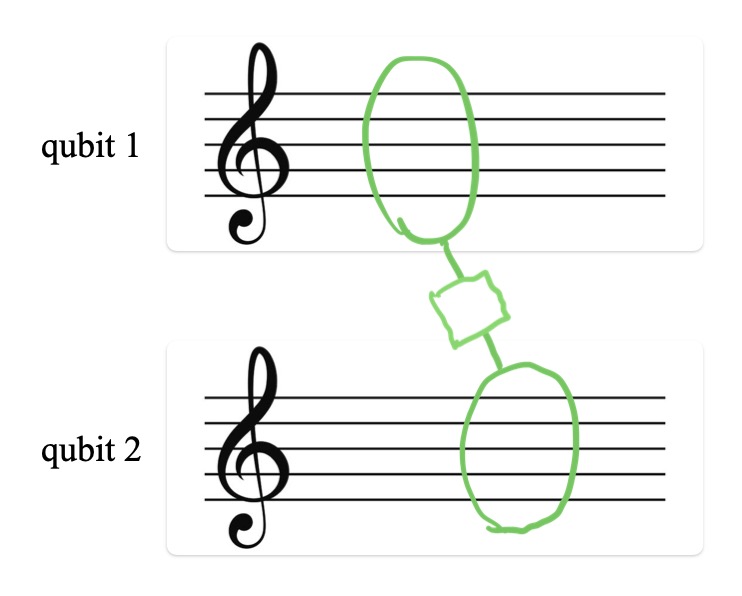,width=100pt}    
\]

\section{``Bell" score}\label{sec:Bell1}

We  restrict ourselves to two qubits under measurement, given some prior entanglement that for simplicity we  take to be fixed for the whole score.  The measurement specifics, including who/what decides when and which measurement takes place is to be specified in the meta-glossary.  This is the composition  ``Bell", made up from some movements that can be specified in the meta-glossary:
\begin{center}
\fbox{
\begin{tabular}{cc}
\epsfig{figure=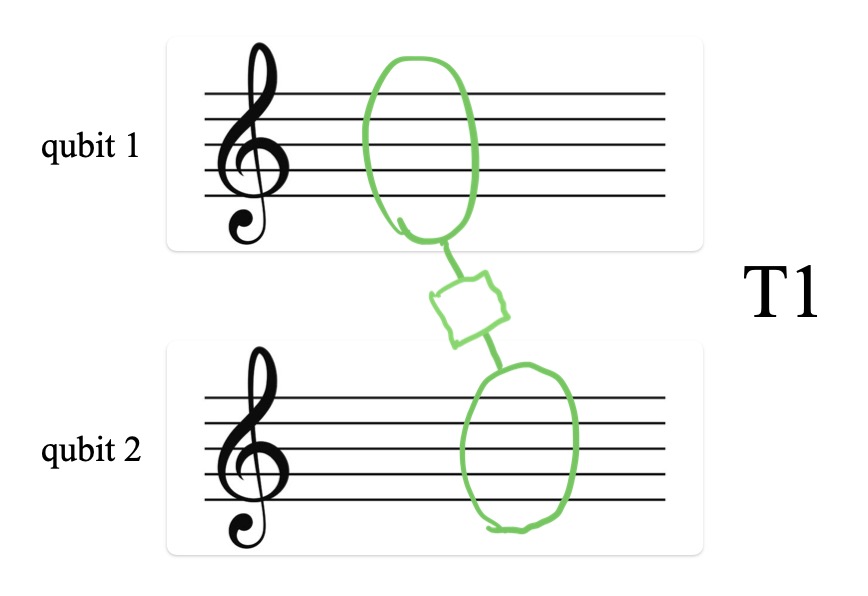,width=110pt}    &\epsfig{figure=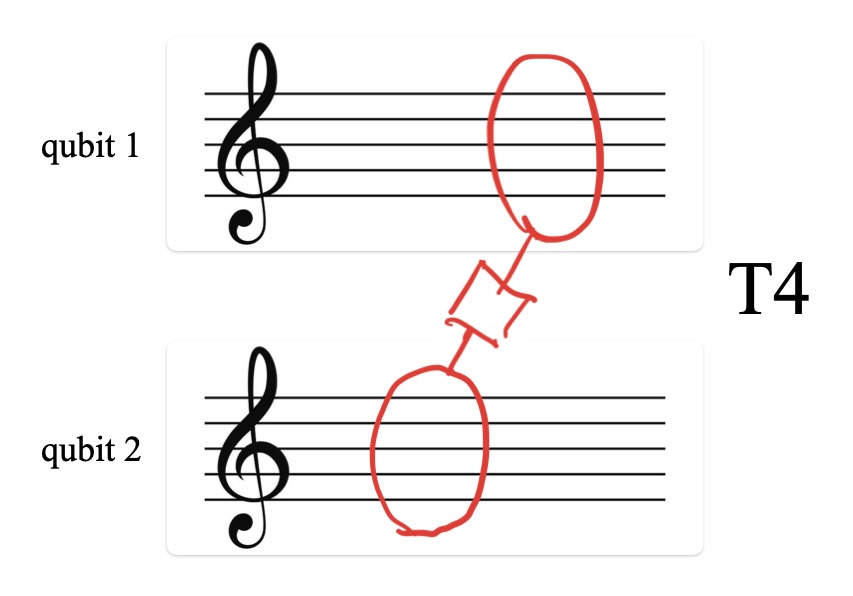,width=110pt}    \\ 
\epsfig{figure=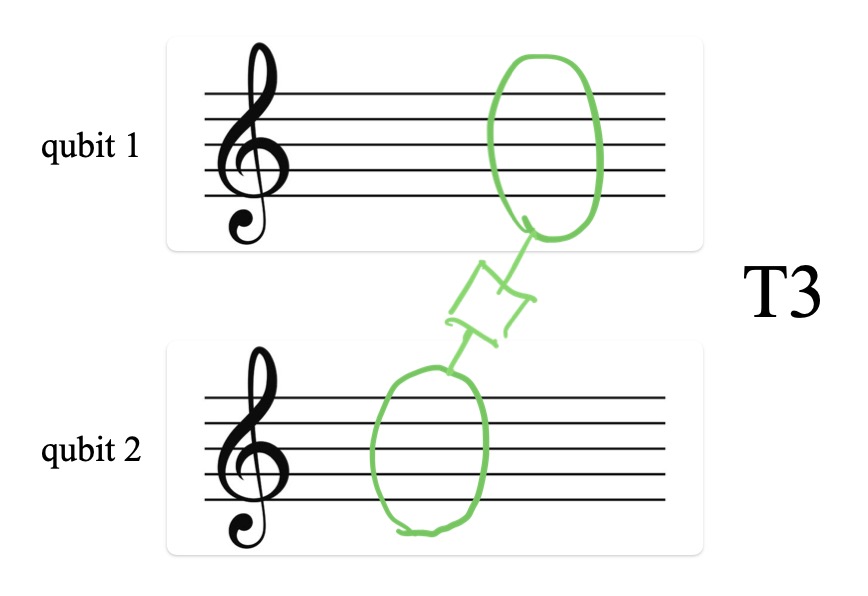,width=110pt}    &\epsfig{figure=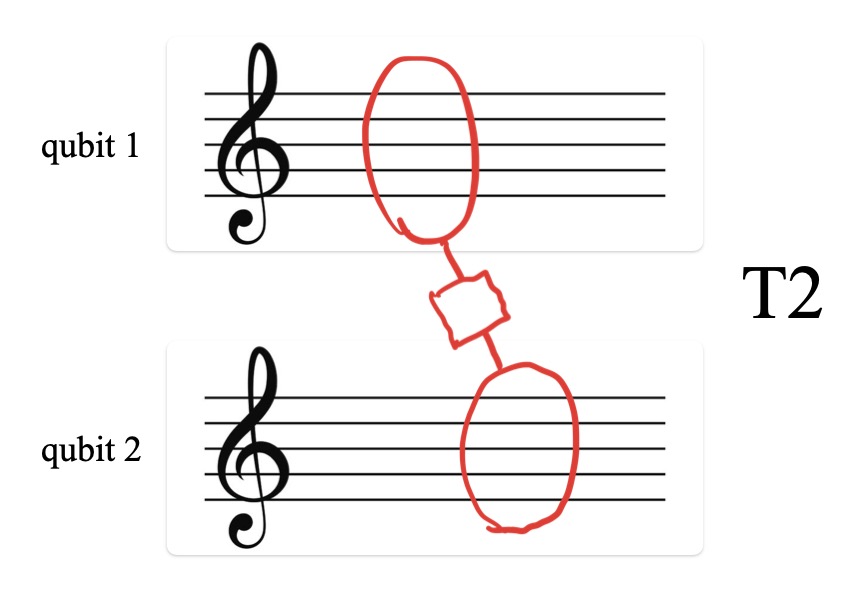,width=110pt}
\end{tabular}
}    
\end{center}
Note that we put a fixed gate, which is unavoidable when the two qubits are realised in a different manner,  and hence need a means of identifying states and measurements.  We illustrate this on an example below.


\section{``Bell" example arrangement}\label{sec:Bell1}

The arrangement described here has already been performed--see Fig.~\ref{fig:Berlin}. The core idea of this quantum-musical experiment was to demonstrate the applicability of Quantum Concept Music to music performance.  From a musical perspective, seen as a musical experiment, it adopts the following \bM features\e: 
\bit\item 
Performance in music  can be conducted using an analogue of quantum measurement and entanglement, in particular, musicians can interact as an entangled pair under measurements in varying bases.
\item
Live improvisation is a very suitable setting for realizing the above.
\item
The audience can act as both  the observer  and instigator of empirical changes. 
\eit
  
  Concretely, the arrangement for the ``Bell" score 
involves the `performance' of two qubits--say qubit 1 and qubit 2--on two different instruments, that will be forced to interact in a very particular way.  

\subsection{Qubit  1: Quantum Guitar}\label{sec:qgit}

One example of a qubit is  realised by means of the Quantum Guitar introduced in \cite{quantumguitar}.   The realisation of the qubit is very direct here, as quantisation of a guitar constitutes associating a qubit to any of its playable states, using Moth's Actias  synth 
of Figure \ref{ActiasQubit}--heavily inspired by the earlier Q1Synth \cite{miranda2023q1synth}. 

\begin{figure}[h!]
   \centerline{\includegraphics[height=120pt]{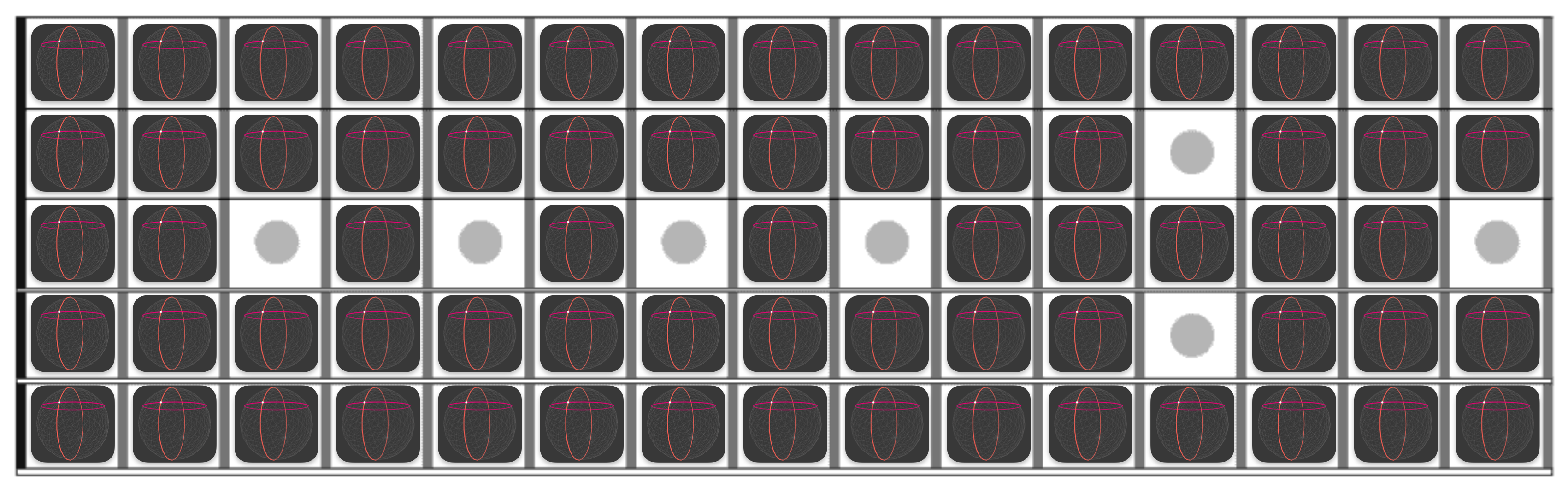}} 
   \caption{\bM Following Deutsch, a quantum field can be seen as identically-constituted quantum physical systems on spacetime, where each event is associated with one such system--and the simplest possible such system is a qubit.  Here space is the fretboard.\e} 
    \label{fig:Fretboard}
\end{figure}

\bM Using a quantum physics analogy, and more specifically, quantum fields, the pitch constitutes the classical  spatial background, which quite literally, is the fretboard of the guitar. The idea of viewing a quantum field by associating a qubit to each spatial location is based on papers by Deutsch and others \cite{Deutsch2004,Pienaar2011,Singh2019}. Deutsch considers a field of identically-constituted quantum physical systems on spacetime, where each event is associated with one such system--and the simplest possible such system is a qubit. The timbre is the result of the additional quantum degrees of freedom, that are provided by the associated qubit. So effectively, we quantize the fretboard of a guitar--see Fig.~\ref{fig:Fretboard}.  
The additional quantum variables provided by Quantum Guitar apply to all notes that are played, and we can change the associated qubit by changing those variables while changing pitch.  

Controlling the qubit, rotating it and measuring it, is done by means of MIDI foot pedals, in order not to interfere with the hands playing the notes\e.

\subsection{Qubit  2: Grand Piano}\label{sec:pianoqubit} 

\bM Here, a musical qubit,  together with two complementary measurements on it, is mentally created.  Projecting the Bloch-sphere on a plane in the usual way, we  will describe two orthogonal lines in the plane, each spanned by a pair of eigenstates, which then  also define the two complementary measurements:
\bit
\item Soft (Delicate) vs Strong (Loud)
\item Slow vs Fast
\eit
These two sets of opposite states are `generic' in the sense their ambiguity and `wideness' of interpretations allows the performers to precisely catch each other's improvisation. This is of a crucial importance for an unprepared improvisation. In addition, musical interpretations enrich the meaning of these states in the following
way:\e
\bit 
\item Soft can mean: tender, delicate, quiet (dynamics), calm (character), etc.
\item Strong can refer to: loud (dynamics), energetic (character), the density of sounds
(e.g.,~many-tone cluster-like chords), the nature of effective passages, etc.
\item  Slow obviously refers to tempo, rhythmic formulas, character, etc.
\item Fast also obviously refers to tempo, rhythmic formulas, character, etc.
\eit
Instead of focusing on specific rhythmic formulas and/or exact pitch-based quasi-melodic formulations, we focused on generic concepts and notions of music. This was crucial because we improvise live, without any preparations, without any pre-determined plan. We can start at a certain point, and arrive later to another point,
but the trajectory of how we move is undetermined. Both for the audience and for us. Even we ourselves are unaware of where this improvisation will lead

\bM These pairs obey a weak analogue quantum complementarity in the sense that being strict in one observable tells nothing about the other one.  An alternative stronger notion of complementarity arises   by considering the following alternative pairs of eigenstates:
\bit
\item (Soft and Slow) vs (Strong and Fast)
\item (Soft and Fast) vs (Strong and Slow)
\eit
As states these are now all four mutually  exclusive.  Moreover, respecting Bloch-sphere geometry, there is now no unique path between the two eigenstates, for example, one can go from (Soft and Slow) to (Strong and Fast) equally well passing either via (Soft and Fast) or via (Strong and Slow).  Moreover, just like on the Bloch-sphere, there are paths to go through neither.  Qubit 2 is depicted in Fig.~\ref{PianoQubit}.  Strictly speaking, here we are only filling half of the sphere, which corresponds, in ZX-calculus terms, for the observables only allowing phases to take values in $[0, \pi]$.  An additional binary variable would fill the full Bloch-sphere.
\e

\subsection{Combining the two}

Another aspect of consideration was that each of us comes from different music
background--classical/avant-garde and industrial music. Thus, referring primarily to
foundational conceptions in music was useful as it allowed us to cross the boundaries
of specific genres and create music on the intersection of diverse styles.

\bM\subsection{Example sniped}\label{sniped}\e

During a performance, we assume that there are two qubits, performing on two instruments, grand piano and quantum guitar respectively. We act as an entangled pair. Any change introduced by one affects the other.

\bM In this sniped, we will use the interpretations of complementarity already indicated above, which on the one hand, affect the Actias-timbre of the guitar, and on the other hand, the tempo and dynamics/character on the piano:
\[
\epsfig{figure=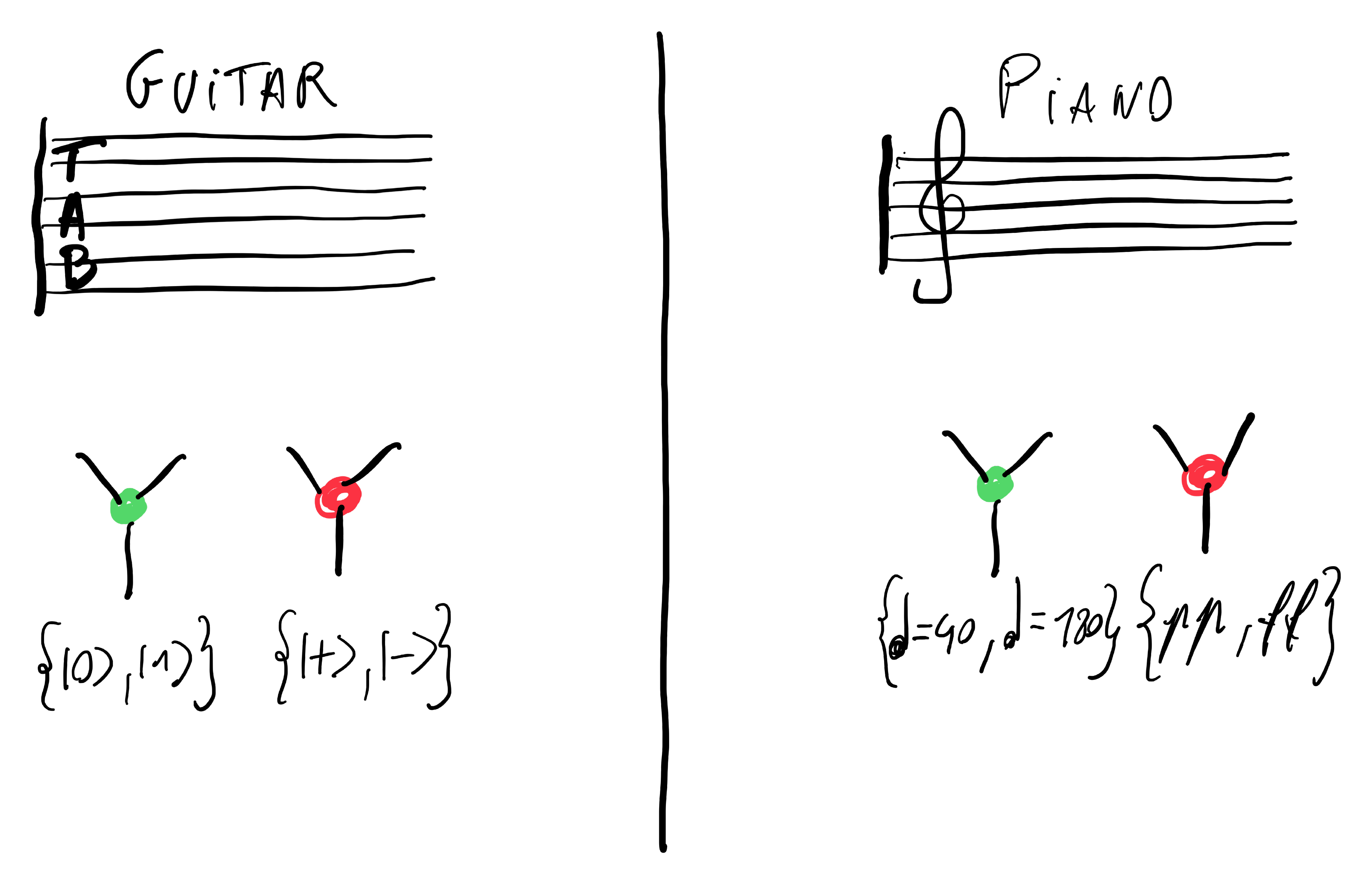,width=240pt}   
\]
We now explain what the following would mean here:
\[
\epsfig{figure=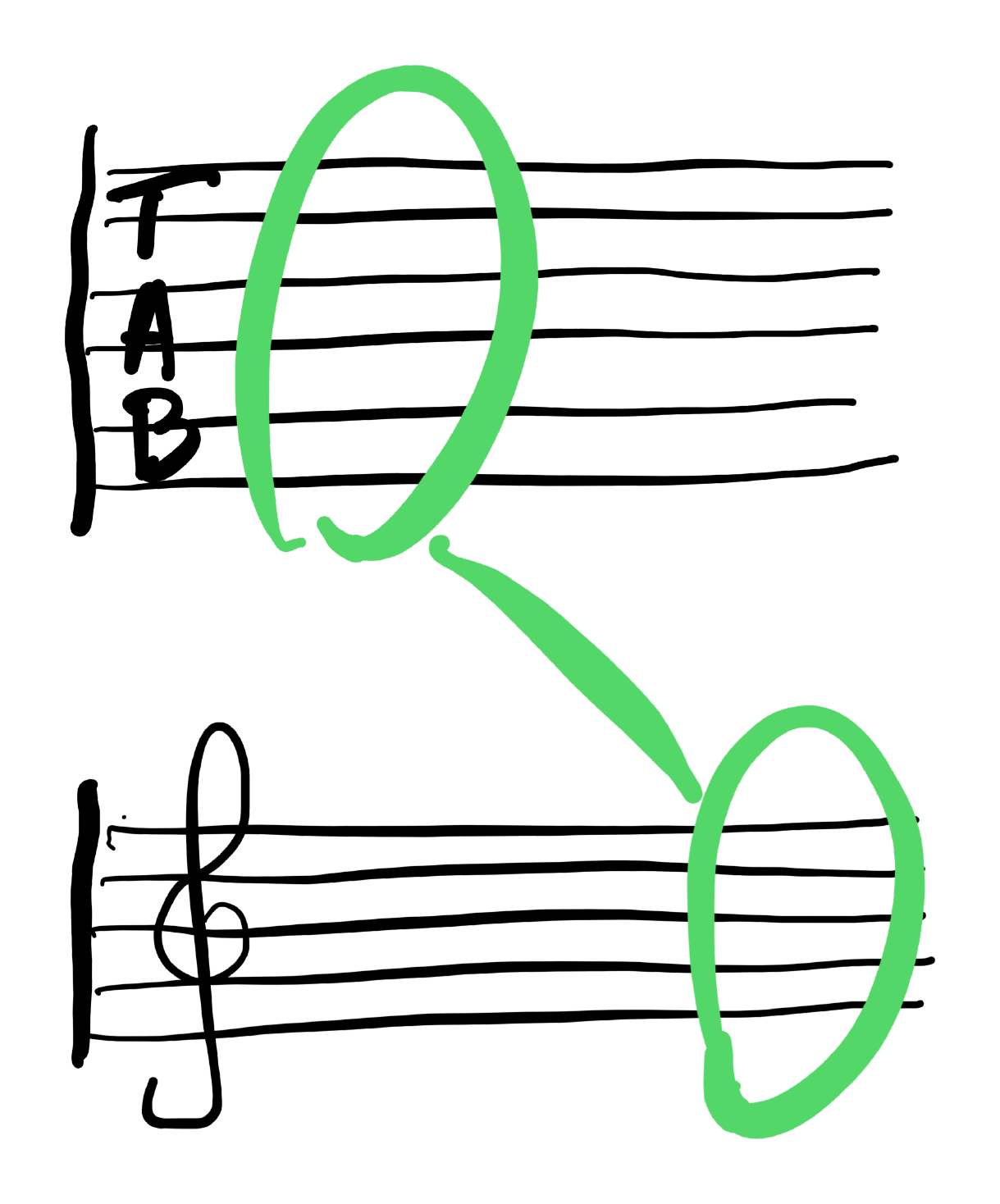,width=65pt}   
\]
Firstly, this choice of complementarity doesn't affect the note-playing at all, just to keep things simple.  Assume that the Actias-timbre state is $|+\rangle$ before, and the piano is playing {\bf Allegro}. At this point a quantum measurement on Actias is initiated, with two possible outcomes, $|0\rangle$ and $|1\rangle$.  Depending on the outcome we either get after the measurement:
\[
\epsfig{figure=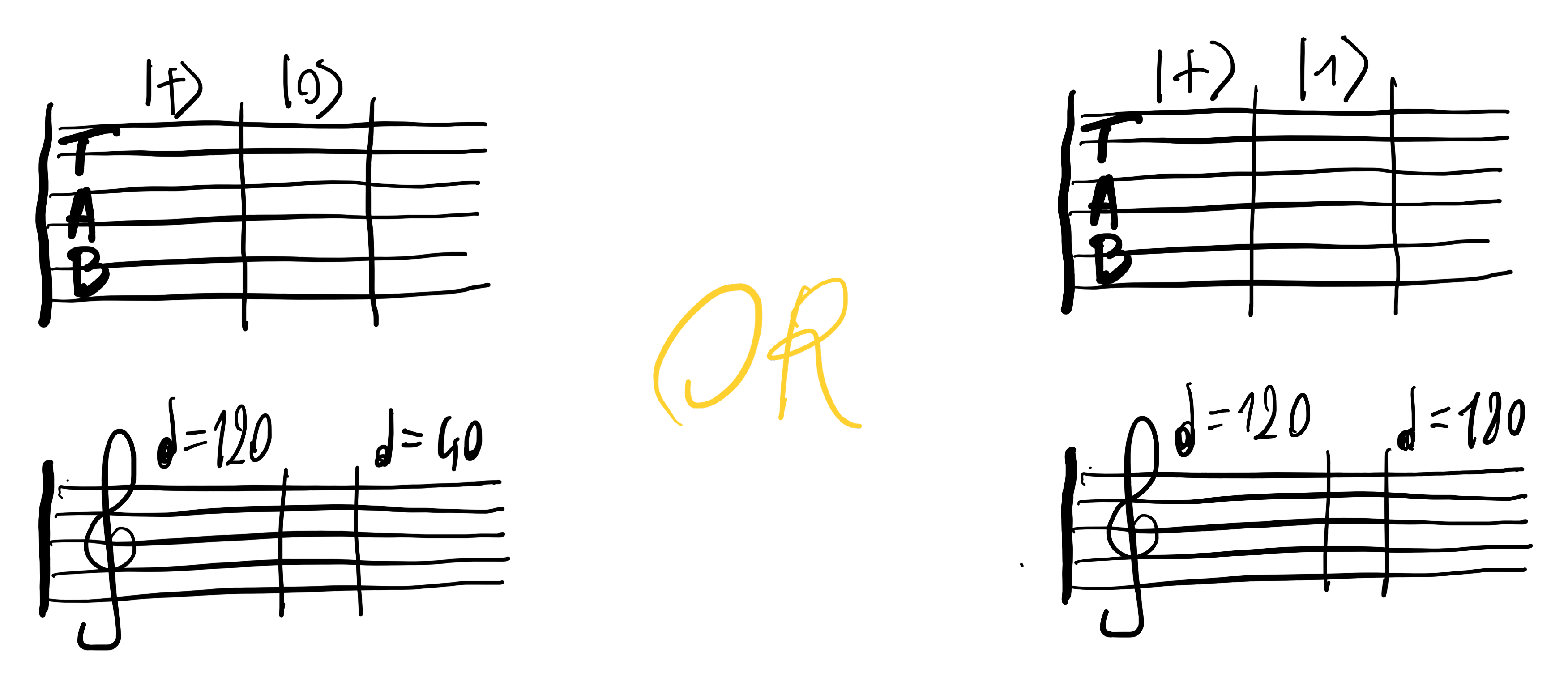,width=210pt}   
\]

Assume now that the Actias-timbre state is $|0\rangle$ before, and the  piano is playing {\bf mezzo forte}. At this point measurement is instigated by the pianist, with two possible outcomes, decided on by the pianist, with two possible outcomes {\em \textbf{pp}} and {\em \textbf{ff}}.  We now consider the following:
\[
\epsfig{figure=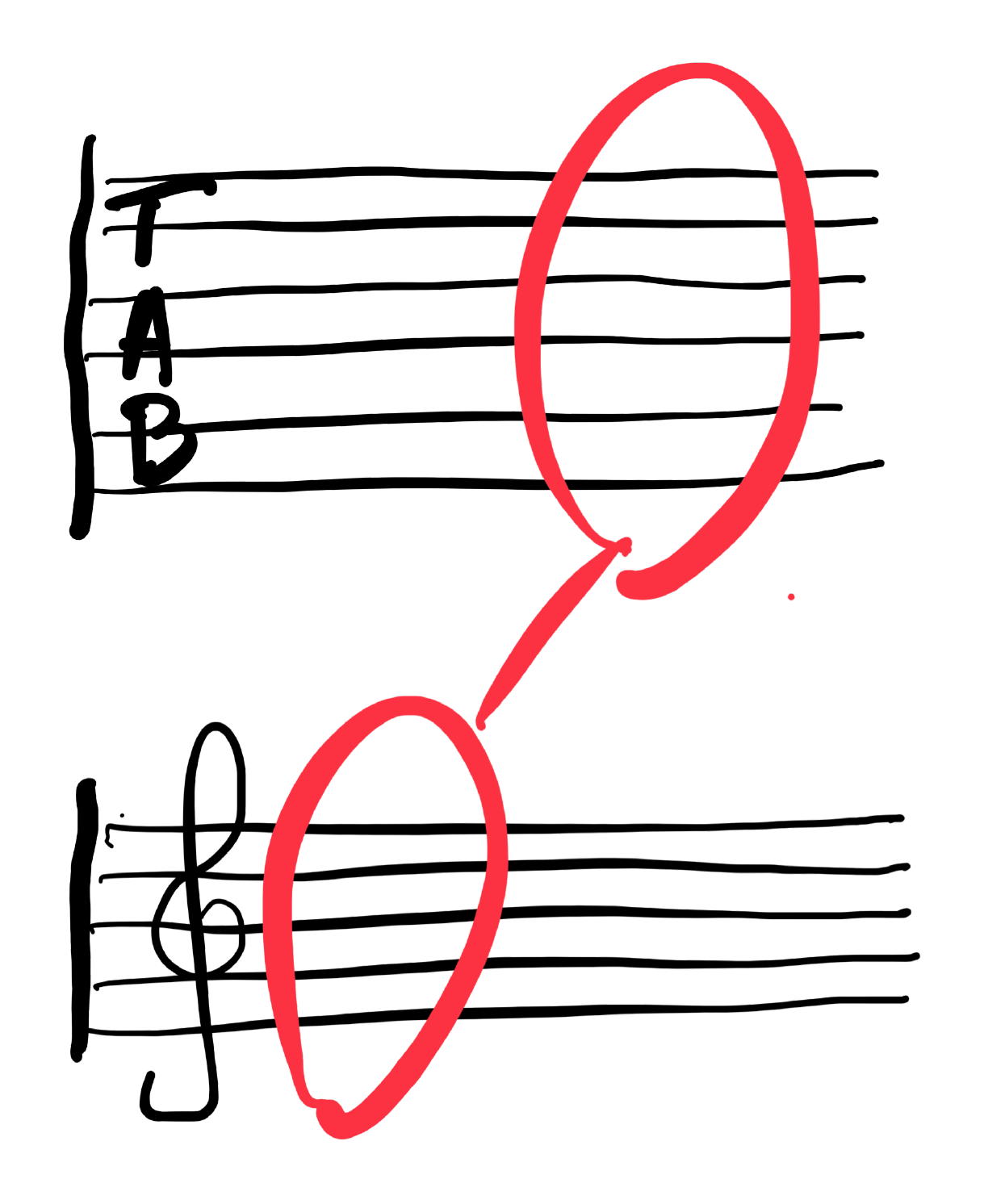,width=65pt}   
\]
Depending on the decided measurement outcome we either get after the measurement:
\[
\epsfig{figure=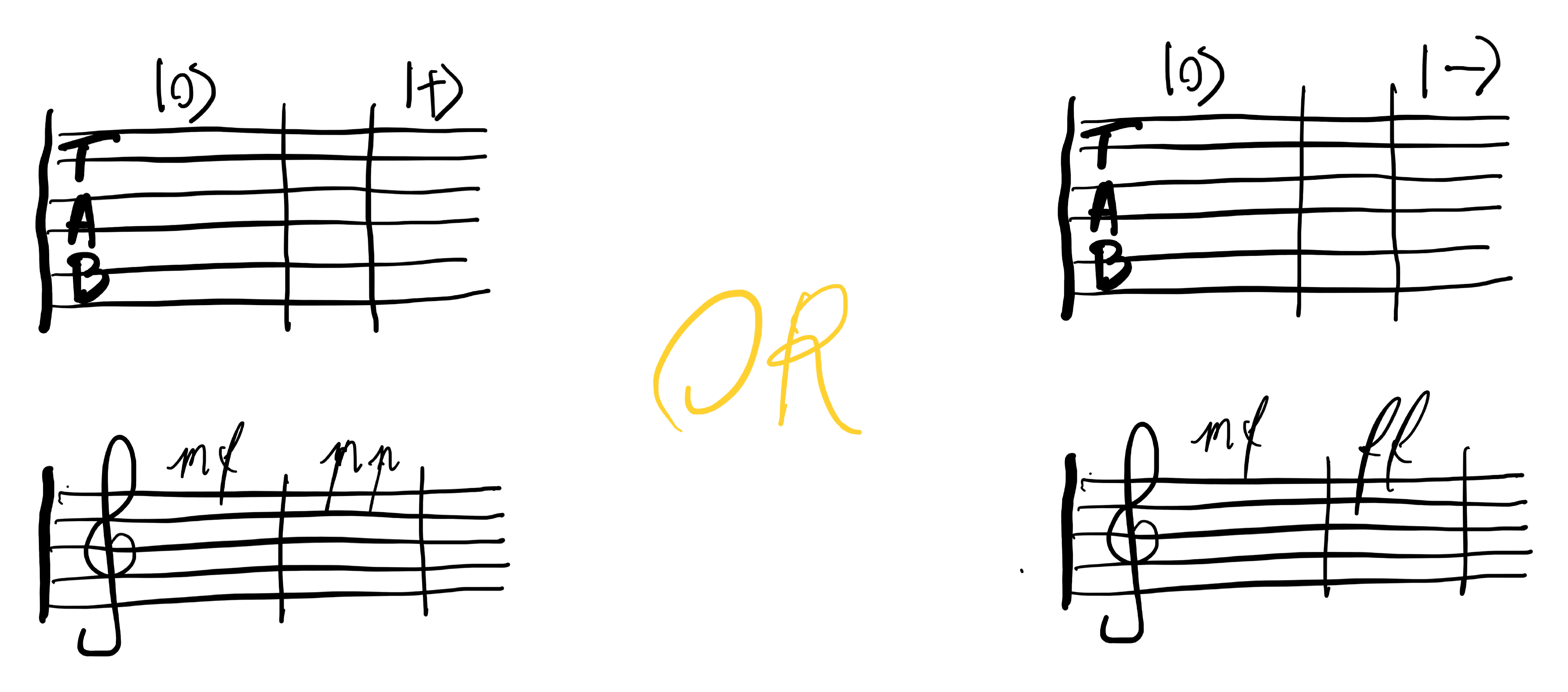,width=210pt}   
\]
We have played this performance at a number of occasions now, it will be elaborated on, and improved.  But it should be clear how how this notation involves the key quantum features: changes of state, non-determinism, caused by measurements.

As a third example, with similar assumptions as above, now using an H-gate:
\[
\epsfig{figure=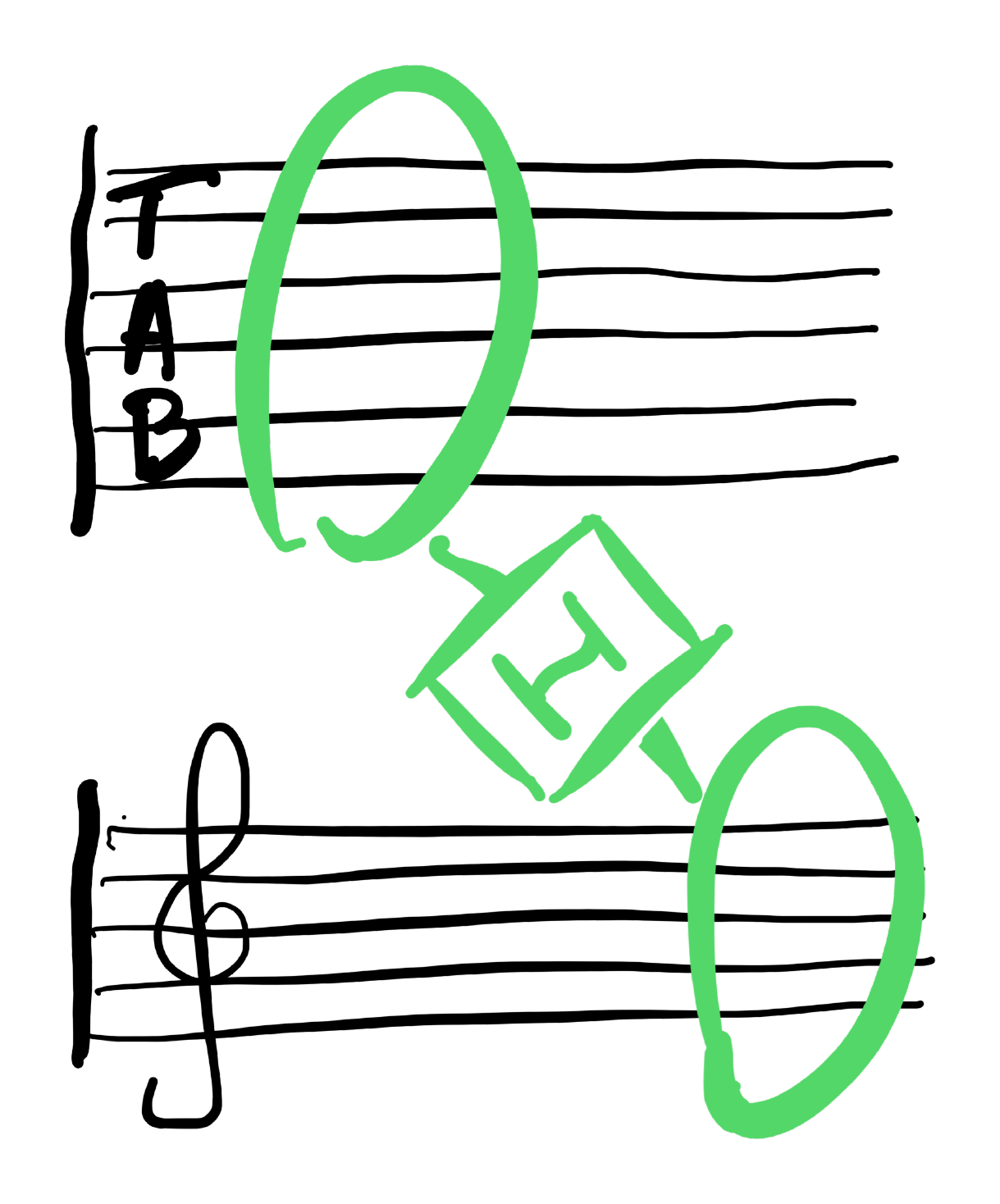,width=65pt}   
\]
we get:
\[
\epsfig{figure=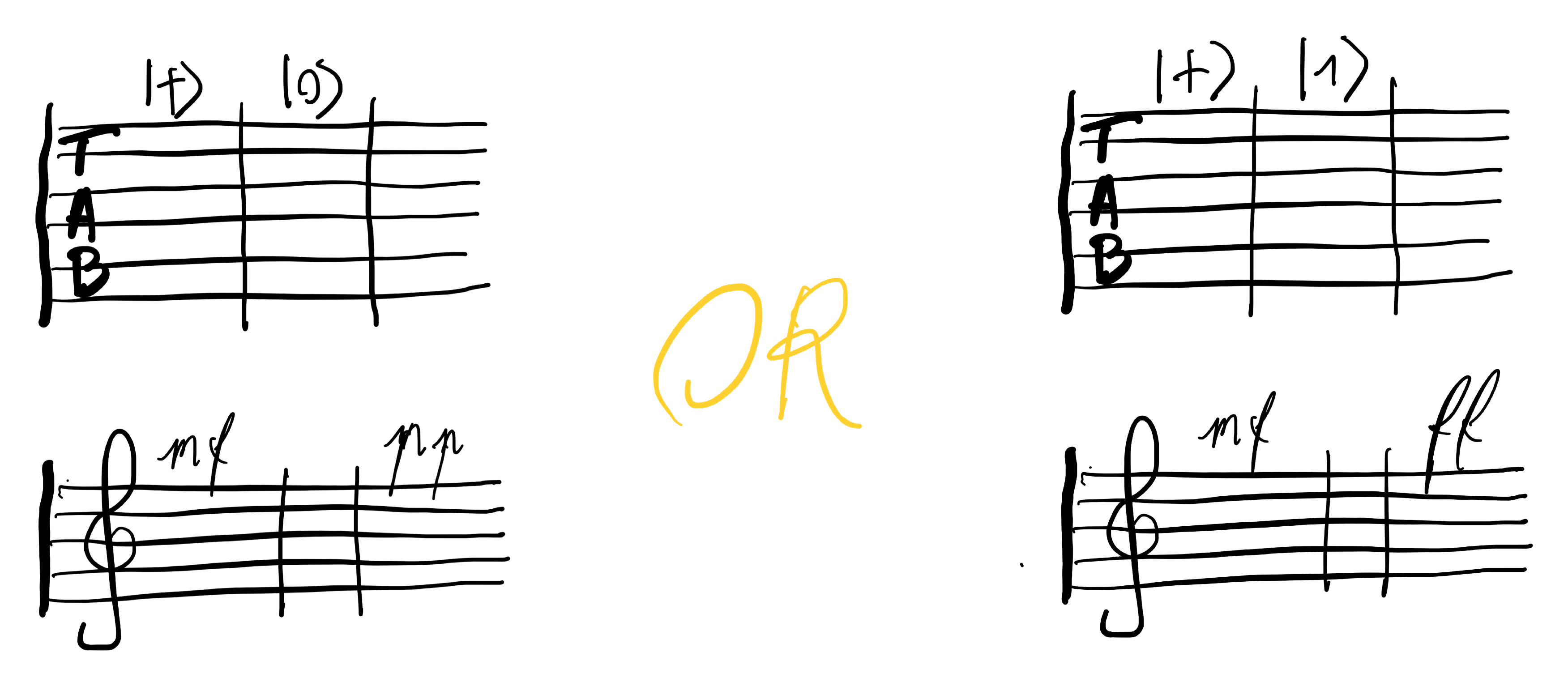,width=210pt}   
\]
given that the H-gate will now map the green basis of the piano on the red basis, resulting in a change of dynamics.
\e






\section{Conflict of interest and code availability statement}

We used software of Moth, of which we are a co-founding shareholder.  It is used by my own band, Black Tish (under QBA), and Moth helped with some concerts (see below).  The open-source software will be redeveloped by QEB and possibly others. The new versions will encompass the current version/data. A previous version of this paper is \cite{ISQCMC_2025_abdyssagin}.

\bM
\section{conclusion}

Here we introduced the beginnings of a quantum-inspired notation for music, that could either be interpreted on traditional instruments, or using quantum systems, e.g.,~via Actias, or a future software that can also handle entangled quantum system.  Such a software is currently being developed by the company Quantum Entertainment Berlin.  There is of course never a conclusion to art. We are part of the process of moving forward, and sometimes moving backward, and then forward again, but mostly, in parallel.

\e

\bibliographystyle{unsrt}
\nocite{miranda2020creative}
\bibliography{mainNOWcopy}

\end{document}